\documentclass[11pt]{article}
\usepackage{graphicx}
\usepackage{latexsym,graphicx}
\usepackage{float}
\usepackage{amsmath}
\usepackage{amsfonts}
\usepackage{amssymb}
\usepackage{hyperref}
\usepackage{epstopdf}
\usepackage{color}
\usepackage{cite}
\usepackage{orcidlink}
\DeclareGraphicsExtensions{.eps}
\usepackage[left=2.5cm,top=2.5cm,right=2.5cm,bottom=2.5cm]{geometry}
\catcode `\@=11
\catcode `\@=12
\begin{document}
	\begin{center}
		\large{\bf{Quintessence dark energy model in non-linear $f(Q)$ theory with bulk-viscosity}} \\
		\vspace{5mm}
		\normalsize{ Dinesh Chandra Maurya}\\
\vspace{5mm}
\normalsize{Centre for Cosmology, Astrophysics and Space Science, GLA University, Mathura-281 406,
	Uttar Pradesh, India.}\\
\vspace{2mm}
E-mail:dcmaurya563@gmail.com \\
	\end{center}
	\vspace{5mm}
	\begin{abstract}
		In this study, we investigate a locally rotationally symmetric (LRS) Bianchi type-I cosmological model in non-linear form of $f(Q)$ gravity with observational constraints. We solved the modified Einstein's field equations with a viscous fluid source and got a hyperbolic solution. First, we apply MCMC analysis to the cosmic chronometer (CC), Baryon Acoustic Oscillation (BAO) and Pantheon datasets to place observational constraints on the model parameters. Using constrained values of model parameters, we study the behavior of cosmological parameters, such as the Hubble parameter $H$, the deceleration parameter $q$, and the equation of state (EoS) parameter $\omega_{v}$ with the skewness parameter $\delta_{v}$ for the viscous fluid. In addition, we perform the Om diagnostics and statefinder analysis to categorize dark energy models. Also, we studied cosmographic series coefficients to explore the whole evolution of the derived universe model. We estimated the current age of the universe as $t_{0}\approx13.8$ Gyrs. We obtained a quintessential and ever-accelerating model with bulk viscosity fluid.
	\end{abstract}
	\smallskip
	\vspace{5mm}
	{\large{\bf{Keywords:}}} LRS Bianchi type-$I$ universe; non-linear $f(Q)$ gravity; bulk-viscosity; analytic solution; observational constraints.\\
	\vspace{1cm}
	
	PACS number: 98.80-k, 98.80.Jk, 04.50.Kd \\
	\section{Introduction}

     Cosmological measurements in 1998 suggest that the late-time universe undergoes an accelerated expansion due to an almost mystical energy with a large negative pressure called ``Dark Energy" \cite{ref1,ref2,ref3,ref4,ref5}. The equation of state (EoS) parameter $\omega$, which is the ratio of energy density to evenly distributed pressure in space, is commonly used to categorize dark energy. Recent cosmological observations suggest that $\omega<-1/3$ is the required value of the EoS parameter to accelerate the expansion of the universe. Scalar field models with an EoS parameter of $-1<\omega<-1/3$ are the leading choices in this category. These are known as Quintessence field dark energy models \cite{ref6, ref7}, whereas $\omega<-1$ is a phantom field dark energy model \cite{ref8}. Among these scenarios of dark energy models, the phantom field dark energy model has received a lot of interest because of its unique features. The phantom model describes developing dark energy that sustains an exciting future spread, culminating in a finite-time future singularity. We know that the EoS parameter for dark energy is $\omega=-1.084\pm0.063$. This information is based on observations obtained by WMAP9 \cite{ref9} and measurements of $H_{0}$, SNe-Ia, the cosmic microwave background, and BAO. In 2015, the Planck collaboration determined that $\omega=-1.006\pm0.0451$ \cite{ref10}.\\
     
     Recent observations have questioned the validity of general relativity (GR), notwithstanding its effectiveness as a physics theory \cite{ref11}. Perhaps the most striking discovery is the fast expansion of our universe in its early and late stages, which general relativity cannot explain properly. Because theory and observation diverge, many theories other than General Relativity (GR) have been proposed. These theories are known as ``modified gravity" \cite{ref12}. They demonstrated how looking for a feasible alternative extended our understanding of gravity. The $f(R)$-gravity concept, introduced in \cite{ref13,ref14}, is the most basic generalization of general relativity. This method requires replacing the Hilbert-Einstein action Ricci-scalar $R$ with a freely chosen function of $R$. The modified $f(R)$ gravity is widely recognized for demonstrating the evolution of the universe, the cosmological constant $\Lambda$, and its impact on acceleration \cite{ref15,ref16}. In the recent literature, several cosmologists have attempted to explain the cosmic acceleration using modified gravity and alternative gravity theories. This startling theory holds that matter fields have no effect on gravitational interactions. A manifold's affine features can be explained by its geometric properties and curvature \cite{ref17,ref18,ref19,ref20}.\\
     
     Torsion, non-metricity, and curvature are all important aspects of metric space connectivity. Torsion and non-metricity are zero in Einstein's standard General Relativity. The equivalence principle states that gravity has a geometric aspect, thus we must consider the various ways it could have a similar geometry. General relativity can be represented as a flat spacetime with an asymmetric connection metric. Torsional forces control gravitational forces in this teleparallel formulation. Our simplified general relativity model uses non-metricity to describe gravity on a flat, uniform spacetime without curvature, as discussed in sources \cite{ref21,ref22,ref23}. The essential assumptions of this geometrical interpretation ensure the future of modified gravity. For example, changing the scalar values for curvature, torsion, and non-metricity in general relativity formulations to arbitrary functions opens up new possibilities for modified gravity theories. New gravity models, especially those based on $f(T)$ \cite{ref24,ref25,ref26} and $f(R)$ \cite{ref27,ref28,ref29}, are becoming popular. This study will focus on the less well-known $f(Q)$ theories, which were first introduced in \cite{ref22}. A recent research by J. Baltran et al. focuses on cosmological topics in $f(Q)$ geometry \cite{ref30}. Harko et al. \cite{ref31} used a power-law function to study matter coupling in $f(Q)$ gravity, and a wide review on $f(Q)$ gravity is given in \cite{ref32}. A recent study \cite{ref33} discovered that the $\Lambda$CDM model may be represented by the equation $f(Q)=Q+\alpha$, where $\alpha$ is a positive value. In the early universe, strings had more mass than particles, but large strings eventually took over. Our latest work have presented the string cosmological model with a constant equation of state parameter in $f(Q)$ gravity theory, as reported in \cite{ref34,ref35,ref36}. Recently, some dark energy models in $f(Q)$ theory with $\Lambda$CDM are well discussed in \cite{n1,n2,n3,n4,n5}\\
     
    Several studies imply that viscous fluids with both shear and bulk viscosity may have contributed to the evolution of the universe \cite{ref37,ref38}. In \cite{ref39,ref40}, parabolic differential equations were employed to explore viscous fluids in relativity. However, they only studied at the first level of deviation from equilibrium. These equations show that heat flow and viscosity spread infinitely, which contradicts particle causality. The concept of second-order divergence from equilibrium was introduced in \cite{ref41,ref42,ref43,ref44} and used to characterize the history of the early universe. A viscous fluid's profligacy process is usually described by its bulk viscosity parameter $\xi$, while its shear viscosity parameter $\eta$ is ignored \cite{ref45,ref46}. Bulk viscosity indicates profligacy. We use the effective pressure $p-3\xi H$ to explain it. Assume $p$ represents isotropic pressure, $\xi$ the bulk viscosity coefficient, and $H$ the Hubble parameter. Entropy generation is positive when $\xi>0$, as established by the second law of thermodynamics \cite{ref47,ref48}.\\
     
     In \cite{ref49,ref50,ref51,ref52}, the influence of bulk viscosity fluid in the late-time accelerated universe was examined. However, in an expanding universe, the viscous fluid has a challenge in establishing a credible mechanism for its creation. Recently, \cite{ref53} has discussed the origin of bulk viscosity in cosmology and its thermodynamical implications. In a theoretical research, the bulk viscosity evolves when the local thermodynamic equilibrium is broken \cite{ref54}. We can think of the bulk viscosity as an effective pressure that returns the system to thermal equilibrium. The bulk viscosity pressure occurs when the cosmic fluid expands or contracts too quickly (i.e., the state deviates from the local thermodynamic equilibrium) \cite{ref55,ref56,ref57} and ends when the fluid regains thermal equilibrium. A viscous cosmology in early an late-time universe is discussed in \cite{ref58} while \cite{ref59} has explored the dynamical properties of the universe using power-law and logarithmic corrected Ricci viscous cosmology. A viscous cosmology with holographic dark energy is discussed in \cite{ref60}. We have recently examined bulk viscosity in a flat and homogeneous universe \cite{ref61,ref62,ref63,ref64} and transit phase universe in $f(Q, T)$ gravity \cite{ref65}.\\
     
     Cosmography has recently attracted significant attention among rational techniques \cite{ref66,ref67,ref68,ref69}. This model-independent method is based purely on the observational assumptions of the cosmological principle, allowing for the study of dark energy evolution without the requirement to use a specific cosmological model. The standard space flight approach is based on Taylor's expansion of observations that may be directly compared to data, and the outcomes of such operations are independent of the state equations used to investigate the universe's evolution. Cosmography is a strong technique for understanding the mechanics of the cosmos \cite{ref70,ref71,ref72,ref73,ref74}.  The cosmological principle specifies a scale factor as the only degree of freedom that rules the universe. By expanding the current Taylor series of $a(t)$ around present time, we can construct the cosmographic series coefficients such as Hubble parameter $(H)$, deceleration parameter $(q)$, jerk $(j)$, snap $(s)$, lerk $(l)$, and max-out $(m)$ as presented in \cite{ref66}.:
     \begin{equation}\nonumber
     	H=\frac{1}{a}\frac{da}{dt},~~q=-\frac{1}{aH^{2}}\frac{d^{2}a}{dt^{2}},~~j=\frac{1}{aH^{3}}\frac{d^{3}a}{dt^{3}}
     \end{equation}
     and
     \begin{equation}\nonumber
     	s=\frac{1}{aH^{4}}\frac{d^{4}a}{dt^{4}},~~l=\frac{1}{aH^{5}}\frac{d^{5}a}{dt^{5}},~~m=\frac{1}{aH^{6}}\frac{d^{6}a}{dt^{6}}
     \end{equation}
     Through the study of these quantities, the dynamics of the late universe are investigated. It is possible to determine the physical features of the coefficients by using the shape of the Hubble expansion when doing so. To be more specific, the sign of the parameter $q$ tells us whether the universe is speeding up or slowing down. It is the sign of $j$ that determines how the dynamics of the universe change, and positive values of $j$ indicate the occurrence of transitional intervals when the expansion phase of the universe changes. It is also necessary for us to have the value of $s$ in order to differentiate between the development of hypotheses regarding dark energy and the behavior of cosmological constant factors.\\
     
     Based on prior research and findings, we study $f(Q)$ gravity in an anisotropic background and solve the field equations for the average scale factor $a(t)$, which is commonly assumed in previous works. We used this scale factor to investigate physical factors, the age of the universe, cosmographic parameters, and the statefinder study of the viscus universe. Section 1 introduces and examines the literature, while Section 2 presents the $f(Q)$ gravity formalism and field equation for LRS Bianchi type I space-time. In Section 3, we solved modified Einstein's field equations using the bulk viscosity factor $\xi(t)=\xi_{1}\dot{H}-\xi_{0}$. In Section 4, we imposed observational limitations on model parameters, and Section 5 investigates the model's physical and kinematic characteristics. The final conclusions are given in Section 6.
     
	\section{Modified Einstein's Field Equations}
		
    As stated in \cite{ref30}, we consider the following action for investigating the universe model in $f(Q)$ gravity:
	\begin{equation}\label{eq1}
		S=\int{d^{4}x~\sqrt{-g}\left[-\frac{1}{2}f(Q)+L_{m}\right]}.
	\end{equation}
    $f(Q)$ denotes any function of the non-metricity scalar  $Q$, $L_{m}$ is the matter Lagrangian, and $g$ is the determinant of the metric tensor $g_{\mu\nu}$. The non-metricity scalar is defined as    
    \begin{equation} \label{eq2}
    	Q\equiv
    	-g^{\mu\nu}(L_{\,\,\,\,\beta\mu}^{\alpha}L_{\,\,\,\,\nu\alpha}^{\beta}-L_{\,\,\,\,\beta\alpha}^{
    		\alpha}L_{\,\,\,\,\mu\nu}^{\beta}).
    \end{equation}
    where $L^{\alpha}_{\,\,\,\,\beta\gamma}$ is the deformation tensor given by, 
    \begin{equation}  \label{eq3}
    	L^{\alpha}_{\,\,\,\,\beta\gamma}=-\frac{1}{2}g^{\alpha\lambda}(\nabla_{\gamma}g_{%
    		\beta\lambda}+\nabla_{\beta}g_{\lambda\gamma}-\nabla_{\lambda}g_{\beta
    		\gamma}).
    \end{equation}
    We define the trace of the non-metricity tensor as
    \begin{equation}\label{eq4}
    	 Q_{\alpha}=g^{\mu\nu}Q_{\alpha\mu\nu},\,\,\,\,\tilde{Q_{\alpha}}=g^{\mu\nu}Q_{\mu\alpha\nu}
    \end{equation}
    We also introduce the superpotential of our model, defined as
    \begin{equation}\label{eq5}
    	{P^{\alpha}}_{\mu\nu}=-\frac{1}{2}{L^{\alpha}}_{\mu\nu}+\frac{1}{4}(Q^{\alpha}-\tilde{Q^{\alpha}})g_{\mu\nu}-\frac{1}{4}\delta^{\alpha}_{(\mu Q_{\nu})}
    \end{equation}
    with the relation    
	\begin{eqnarray}
		Q&=&-Q_{\alpha\mu\nu}P^{\alpha\mu\nu},\\\label{eq6}
		&=&-\frac{1}{4}\left(-Q_{\alpha\mu\nu}Q^{\alpha\mu\nu}+2Q_{\alpha\mu\nu}Q^{\mu\alpha\nu}+Q_{\alpha}Q^{\alpha}-2Q_{\alpha}\tilde{Q}^{\alpha}\right),\\\label{eq7}
		&=&-\frac{1}{4}\left[\nabla_{\alpha}g_{\mu\nu}\nabla^{\alpha}g^{\mu\nu}-2\nabla_{\alpha}g_{\mu\nu}\nabla^{\mu}g^{\alpha\nu}+(g_{\rho\mu}\nabla_{\alpha}g^{\rho\mu})(g_{\sigma\nu}\nabla^{\alpha}g^{\sigma\nu})-2(g_{\mu\rho}\nabla_{\alpha}g^{\mu\rho})(\nabla_{\beta}g^{\alpha\beta})\right],\label{eq8}
	\end{eqnarray}
	with non-metricity tensor $Q_{\alpha\mu\nu}=\nabla_{\alpha}g_{\mu\nu}$, $Q^{\alpha\mu\nu}=-\nabla^{\alpha}g^{\mu\nu}$ and its traces $Q_{\alpha}=-g_{\rho\mu}\nabla_{\alpha}g^{\rho\mu}$, $Q^{\alpha}=-g_{\sigma\nu}\nabla^{\alpha}g^{\sigma\nu}$ and $\tilde{Q}_{\alpha}=\nabla^{\beta}g_{\alpha\beta}$, $\tilde{Q}^{\alpha}=\nabla_{\beta}g^{\alpha\beta}$.\\	
   The field equations are obtained by varying the action (\ref{eq1}) with respect to the metric tensor $g_{\mu\nu}$:	
	\begin{equation}\label{eq9}
		\frac{2}{\sqrt{-g}}\nabla_{\alpha}({\sqrt{-g}f_{Q}{P^{\alpha}}_{\mu\nu}})+\frac{1}{2}g_{\mu\nu}f+f_{Q}(P_{\mu\alpha\beta}{Q_{\nu}}^{\alpha\beta}-2Q_{\alpha\beta\mu}{P^{\alpha\beta}}_{\nu})=T_{\mu\nu},
	\end{equation}
	where $f_{Q}=\partial{f}/\partial{Q}$. Raising one index, we can write the above equation in the form of
	\begin{equation}\label{eq10}
		\frac{2}{\sqrt{-g}}\nabla_{\alpha}({\sqrt{-g}f_{Q}{P^{\alpha\mu}}_{\nu}})+\frac{1}{2}\delta^{\mu}_{\nu}f+f_{Q}P^{\mu\alpha\beta}Q_{\nu\alpha\beta}=T^{\mu}_{\nu}.
	\end{equation}
    The connection is torsion-free, and in the area where we've employed it, the motion connection equation can be easily derived as follows: $\delta_{\xi}{\Gamma^{\alpha}}_{\mu\beta}=-L_{\xi}{\Gamma^{\alpha}}_{\mu\beta}=-\nabla_{\mu}\nabla_{\beta}\xi^{\alpha}$.  In the absence of hypermomentum, the connection field equations have the following form, as the connection's variation with respect to $\xi^{\alpha}$ is homological.	
	\begin{equation}\label{eq11}
		\nabla_{\mu}\nabla_{\nu}(\sqrt{-g}f_{Q}{P^{\mu\nu}}_{\alpha})=0.
	\end{equation}	
    The metric and connection equations can be used to argue that $D_{\mu}{T^{\mu}}_{\nu}=0$, where $D_{\mu}$ is the metric-covariant derivative \cite{ref75}, as it should be due to diffeomorphism invariance. According to reference \cite{ref31}, divergence of the stress-energy-momentum tensor (SEMT) and the hypermomentum indicates a nontrivial hypermomentum.\\
   The SEMT $T_{\mu\nu}$ is expressed as
  \begin{equation}\label{eq12}
  T_{\mu\nu}=-\frac{2}{\sqrt{-g}}\frac{\delta (\sqrt{-g}L_{m})}{\delta g^{\mu\nu}}.
  \end{equation}
  In this work, we looked at the LRS Binachi Type-I spacetime metric element, written as
  \begin{equation}\label{eq13}
  ds^{2}=-dt^{2}+A(t)^{2}dx^{2}+B(t)^{2}(dy^{2}+dz^{2}),
  \end{equation}
  The metric potentials $A(t)$ and $B(t)$ are only functions of cosmic time $t$. The equivalent non-metricity scalar $Q$ is derived as
 \begin{equation}\label{eq14}
  Q=-2\left(\frac{\dot{B}}{B}\right)^{2}-4\frac{\dot{A}}{A}\frac{\dot{B}}{B}.
  \end{equation}
  The SEMT for bulk viscous fluid is taken as
  \begin{equation}\label{eq15}
  T^{\mu}_{\nu}=diag[-\rho, \tilde{p_{x}}, \tilde{p_{y}}, \tilde{p_{z}}],
  \end{equation}
  where $\rho$ denotes the energy density, and $\tilde{p_{x}}$, $\tilde{p_{y}}$, and $\tilde{p_{z}}$ represent the pressures of a viscous fluid occupying the universe along the $x$, $y$, and $z$ axes, respectively.  Taking into account the pressure anisotropy and the equation of state (EoS) parameter, we have
\begin{equation}\label{eq16}
  T^{\mu}_{\nu}=diag[-1, \tilde{\omega_{x}}, \tilde{\omega_{y}}, \tilde{\omega_{z}}]\rho=[-1, \omega_{v}, \omega_{v}+\delta_{v}, \omega_{v}+\delta_{v}]\rho,
\end{equation}
 where $\delta$ is the skewness parameter, indicating the deviation from $\omega_{v}$ along the $y$ and $z$ axes ($\tilde{\omega_{x}}=\omega_{v}$). The parameters $\omega_{v}$ and $\delta_{v}$ are variable and may depend on cosmic time $t$. Utilizing a co-moving coordinate system, we can resolve the field equations \eqref{eq10} for the metric specified in \eqref{eq13} as below:
 \begin{eqnarray}
 	f_{Q}\left[4\frac{\dot{A}}{A}.\frac{\dot{B}}{B}+2\left(\frac{\dot{B}}{B}\right)^{2}\right]-\frac{f}{2}&=&\rho,\label{eq17}\\
 	2f_{Q}\left[\frac{\dot{A}}{A}.\frac{\dot{B}}{B}+\frac{\ddot{B}}{B}+\left(\frac{\dot{B}}{B}\right)^{2}\right]-\frac{f}{2}+
 	2\frac{\dot{B}}{B}\dot{Q}f_{QQ}&=&-\tilde{p_{x}},\label{eq18}\\
 	f_{Q}\left[3\frac{\dot{A}}{A}.\frac{\dot{B}}{B}+\frac{\ddot{A}}{A}+\frac{\ddot{B}}{B}+
 	\left(\frac{\dot{B}}{B}\right)^{2}\right]-\frac{f}{2}+\left(\frac{\dot{A}}{A}+\frac{\dot{B}}{B}\right)\dot{Q}f_{QQ}&=&-\tilde{p_{y}}=-\tilde{p_{z}},\label{eq19}
 \end{eqnarray}
where the dot (.) signifies the derivative concerning cosmic time $t$.\\
The spatial volume for the LRS Bianchi type-I model is expressed as
\begin{equation}\label{eq20}
  V=a(t)^{3}=AB^{2},
\end{equation}
$a(t)$ represents the Universe's average scale factor. The deceleration parameter $(q)$ is defined as:
\begin{equation}\label{eq21}
  q=-\frac{a\ddot{a}}{\dot{a}^{2}}.
\end{equation}
  The deceleration parameter $(q)$ reveals the evolution phase of the expanding universe. The parameter $q$ is positive $(q > 0)$ when the universe experiences deceleration over time and negative $(q < 0)$ in the context of an accelerating universe. The average Hubble parameter, denoted as $H$, is defined as
\begin{equation}\label{eq22}
  H=\frac{1}{3}(H_{x}+H_{y}+H_{z}),
\end{equation}
  Here, $H_{x}$, $H_{y}$, and $H_{z}$ represent the directional Hubble parameters along the $x$, $y$, and $z$ axes, respectively.  According to Eq.\,\eqref{eq13}, the parameters are expressed as $H_{x}=\frac{\dot{A}}{A}$ and $H_{y}=H_{z}=\frac{\dot{B}}{B}$.\\
 The Hubble parameter, spatial volume, and average scale factor are interrelated.
\begin{equation}\label{eq23}
  H=\frac{1}{3}\frac{\dot{V}}{V}=\frac{1}{3}\left[\frac{\dot{A}}{A}+2\frac{\dot{B}}{B}\right]=\frac{\dot{a}}{a}.
\end{equation}
The scalar expansion $\theta(t)$, shear scalar $\sigma^{2}(t)$, and the mean anisotropy parameter $\Delta$ are defined as follows:
\begin{equation}\label{eq24}
  \theta(t)=\frac{\dot{A}}{A}+2\frac{\dot{B}}{B},
\end{equation}
\begin{equation}\label{eq25}
  \sigma^{2}(t)=\frac{1}{3}\left(\frac{\dot{A}}{A}-\frac{\dot{B}}{B}\right)^{2},
\end{equation}
\begin{equation}\label{eq26}
  \Delta=\frac{1}{3}\sum_{i=1}^{3}\left(\frac{H_{i}-H}{H}\right)^{2},
\end{equation}
where $H_{i}, i=1, 2, 3$ are directional Hubble parameters.
\section{Solution of the field equations}
The field equations \eqref{eq17}, \eqref{eq18}, and \eqref{eq19} form a system of three independent equations involving six unknowns: $A$, $B$, $f(Q)$, $Q$, $\omega$, and $\delta$. The system is initially indeterminate. Additional physical constraints are necessary to obtain exact solutions for the field equations. Initially, we apply a physical condition where shear is proportional to the expansion scalar $(\sigma\propto\theta)$. This results in the relationship
\begin{equation}\label{eq27}
  A=B^{m},
\end{equation}
where \( m \neq 1 \) is an arbitrary constant. In the case where \( m=1 \), an isotropic model is obtained; in all other instances, the model is anisotropic. Studies on the velocity redshift relation for extragalactic sources \cite{ref76} say that the universe may reach isotropy when $\frac{\sigma}{\theta}$ stays the same. A few cosmologists have also said that for metrics that are uniform in space, normal congruence to the homogeneous expansion gives a value of about $0.3$ for $\frac{\sigma}{\theta}$ \cite{ref77}. From a study of the 4-year CMB data by Bunn et al. \cite{ref78}, we can see that the shear $\left(\frac{\sigma}{H}\right)$ has a high upper limit of less than $10^{-3}$ in the Planck era. Since the Bianchi models show anisotropic space-time, or $\frac{\sigma}{\theta}=l$, where $l$ is a constant, the ratio of the shear and expansion scalars is thought to be constant. Tise condition has been addressed multiple times in the literature \cite{ref79,ref80,ref81}.\\
Utilizing relation \eqref{eq27} in Eq.\,\eqref{eq20}, we derive the metric coefficients as follows:
\begin{equation}\label{eq28}
  A=a(t)^{\frac{3m}{m+2}}, \hspace{1cm} B=a(t)^{\frac{3}{m+2}}.
\end{equation}
The pressure of a viscous fluid is defined in the $x$, $y$, and $z$ directions \cite{ref47}, as
\begin{equation}\label{eq29}
\tilde{p_{x}}=p-3\xi(t)H_{x}\hspace{1cm} \tilde{p_{y}}=p-3\xi(t)H_{y}\hspace{1cm}\tilde{p_{z}}=p-3\xi(t)H_{z}.
\end{equation}
Here, $p$ represents the normal pressure, while $\xi$ is produced in the viscous fluid that deviates from local thermal equilibrium. Additionally, $\xi$ may depend on the Hubble parameter and its derivatives \cite{ref40,ref82}.\\
We consider the non-linear quadratic form of the $f(Q)$ function.
\begin{equation}\label{eq30}
	f(Q)=-\alpha Q^{2},
\end{equation}
where $\alpha$ is an arbitrary constant. This quadratic form of \( f(Q) \) yields the standard field equations of the non-linear \( f(Q) \) theory of gravity that govern the LRS Bianchi type-I Universe.\\
Applying Eq.\,\eqref{eq29} and subtracting \eqref{eq19} from \eqref{eq18} yields
\begin{equation}\label{eq31}
f_{Q}\left[\frac{\dot{A}}{A}\frac{\dot{B}}{B}+\frac{\ddot{A}}{A}-\frac{\ddot{B}}{B}-\left(\frac{\dot{B}}{B}\right)^{2}\right]+\left(\frac{\dot{A}}{A}-\frac{\dot{B}}{B}\right)\dot{Q}f_{QQ}+3\xi(t)(H_{x}-H_{y})=0.
\end{equation}
From Eq.\,\eqref{eq30}, we derive
\begin{equation}\label{eq32}
  f_{Q}=-2\alpha Q, \hspace{1cm} f_{QQ}=-2\alpha,
\end{equation}
and using Eq.\,\eqref{eq28} in \eqref{eq14}, we get the non-metricity scalar as
\begin{equation}\label{eq33}
  Q=-\frac{18(2m+1)}{(m+2)^2}\left(\frac{\dot{a}}{a}\right)^{2}.
\end{equation}
Applying Eq.\,\eqref{eq29} for a viscous universe in Eqs.\,\eqref{eq18} and \eqref{eq19}, we determine that the bulk viscosity coefficient $\xi$ is associated with matter, the Hubble parameter, and its derivative. Thus, we assume $\xi=\xi(H)$ and examine a specific form of $\xi$ as referenced in \cite{ref58,ref83,ref84,ref85,ref86}.
\begin{equation}\label{eq34}
  \xi(t)=\xi_{1}\dot{H}-\xi_{0},
\end{equation}
where $\xi_{0}$ and $\xi_{1}$ are arbitrary constants.\\
From Eqs.\,\eqref{eq27} to \eqref{eq34}, we get
\begin{equation}\label{eq35}
 \dot{H}+\frac{36\alpha(2m+1)}{36\alpha(2m+1)+\xi_{1}(m+2)^{2}}H^{2}-\frac{\xi_{0}(m+2)^{2}}{36\alpha(2m+1)+\xi_{1}(m+2)^{2}}=0.
\end{equation}
Solving Eq.\,\eqref{eq35} for the average Hubble parameter $H(t)$, we get
\begin{equation}\label{eq36}
  H(t)=k_{0}\coth(k_{1}t+c_{0}),
\end{equation}
where $c_{0}$ is an arbitrary constant and $k_{0}=\frac{(m+2)\sqrt{\xi_{0}}}{6\sqrt{\alpha(2m+1)}}$, and $k_{1}=\frac{6(m+2)\sqrt{\alpha\xi_{0}(2m+1)}}{36\alpha(2m+1)+\xi_{1}(m+2)^{2}}$. Again integrating Eq.\,\eqref{eq35} for the scale factor $a(t)$, we obtain
\begin{equation}\label{eq37}
  a(t)=c_{1}[\sinh(k_{1}t+c_{0})]^{n},
\end{equation}
where $c_{1}$ is an integrating constant and $n=\frac{36\alpha(2m+1)+\xi_{1}(m+2)^{2}}{36\alpha(2m+1)}$.\\
Now, using the relationship of scale factor $a(t)$ with redshift $z$, $(1+z)^{-1}=a(t)a_{0}^{-1}$, \cite{ref87} with Eq.\,\eqref{eq37}, we rewrite the Hubble function as
\begin{equation}\label{eq38}
	H(z)=\frac{H_{0}}{\sqrt{1+c_{1}^{\frac{2}{n}}}}\sqrt{1+[c_{1}(1+z)]^{\frac{2}{n}}},
\end{equation}
where $\frac{(m+2)\sqrt{\xi_{0}}}{6\sqrt{\alpha(2m+1)}}=\frac{H_{0}}{\sqrt{1+c_{1}^{\frac{2}{n}}}}$.\\
The deceleration parameter $q(z)$ is obtained as
\begin{equation}\label{eq39}
q(z)=-1+\frac{36\alpha(2m+1)}{36\alpha(2m+1)+\xi_{1}(m+2)^{2}}\frac{[c_{1}(1+z)]^{\frac{2}{n}}}{1+[c_{1}(1+z)]^{\frac{2}{n}}}.
\end{equation}
From the Eqs.\,\eqref{eq18} and \eqref{eq19}, we derive the directional EoS parameter $\omega_{x}$, $\omega_{y}$ and skewness parameter $\delta_{v}$ for viscous fluid, respectively, as
\begin{equation}\label{eq40}
	\omega_{x}=\omega_{v}=-\frac{3(2m+3)}{5(2m+1)}+\frac{144\alpha(m+2)}{5[36\alpha(2m+1)+\xi_{1}(m+2)^{2}]}\frac{[c_{1}(1+z)]^{\frac{2}{n}}}{1+[c_{1}(1+z)]^{\frac{2}{n}}},
\end{equation}
\begin{equation}\label{eq41}
	\omega_{y}=-\frac{2m^{2}+8m+5}{5(2m+1)}+\frac{72\alpha(m+1)(m+2)}{5[36\alpha(2m+1)+\xi_{1}(m+2)^{2}]}\frac{[c_{1}(1+z)]^{\frac{2}{n}}}{1+[c_{1}(1+z)]^{\frac{2}{n}}},
\end{equation}
\begin{equation}\label{eq42}
	\delta_{v}=\frac{2(2-m-m^{2})}{5(2m+1)}+\frac{72\alpha(m+2)(m-1)}{5[36\alpha(2m+1)+\xi_{1}(m+2)^{2}]}\frac{[c_{1}(1+z)]^{\frac{2}{n}}}{1+[c_{1}(1+z)]^{\frac{2}{n}}},
\end{equation}
where $n=\frac{36\alpha(2m+1)+\xi_{1}(m+2)^{2}}{36\alpha(2m+1)}$.
\section{Observational Constraints}

Our research employs Hubble measurements acquired via two principal methodologies and the Pantheon sample of SNe Ia observations. The initial method entails grouping galaxies or quasars, facilitating a direct assessment of the Hubble expansion by detecting the Baryon Acoustic Oscillation (BAO) peak in the radial direction \cite{ref88}. The second way utilizes the differential age technique, known as the cosmic chronometers (CC) method. This approach relies on the correlation between the Hubble parameter and the temporal derivative of the redshift of remote entities, such as substantial elliptical galaxies. The relationship is articulated as $H(z) = -\frac{1}{(1+z)}\frac{dz}{dt}$ \cite{ref89}, facilitating the calculation of $H(z)$ by the measurement of the relative ages of these objects at varying redshifts. Employing emcee software \cite{ref90}, we perform MCMC analysis on the CC, BAO and Pantheon datasets, minimizing $\chi^{2}$ and maximizing $\mathcal{L}\propto{e^{-\chi^{2}}}$ with suitable priors and covariance matrices to restrict cosmological parameters and examine the expansion phase.\\
\begin{table}[!t]
	\centering
	\begin{tabular}{|c|c|c|c|c|c|c|c|}
		\hline
		\multicolumn{8}{|c|}{CC data} \\
		\hline
		\textbf{$z$} & \textbf{$H(z)$} & \textbf{$\sigma_{H}$} & Ref. & \textbf{$z$} & \textbf{$H(z)$} & \textbf{$\sigma_{H}$} & Ref. \\
		\hline
		0.07  & 69.0  & 19.6  & \cite{ref91} &  0.4783 & 83.8  & 10.2 & \cite{ref95}\\
		0.09  & 69.0  & 12.0  & \cite{ref92}  & 0.48  & 97.0  & 62.0  & \cite{ref97} \\
		0.12  & 68.6  & 26.2  & \cite{ref91} &  0.5929 & 107.0 & 15.5 & \cite{ref94} \\
		0.17  & 83.0  & 8.0   & \cite{ref93}   & 0.6797 & 95.0  & 10.5 &  \cite{ref94} \\
		0.1791 & 78.0  & 6.2  & \cite{ref94}& 0.75 &  98.8 & 33.6 & \cite{ref98}\\
		0.1993 & 78.0  & 6.9  & \cite{ref94} & 0.7812 & 96.5  & 12.5 &  \cite{ref94}\\
		0.2  & 72.9  & 29.6  & \cite{ref91}  & 0.8754 & 124.5 & 17.4 &  \cite{ref94} \\
		0.27  & 77.0  & 14.0  & \cite{ref93}  & 0.88  & 90.0  & 40.0  & \cite{ref97}   \\
		0.28  & 88.8  & 36.6  & \cite{ref91}  & 0.90  & 117.0 & 23.0  & \cite{ref93}  \\
		0.3519 & 85.5  & 15.7 & \cite{ref94} &  1.037  & 133.5 & 17.6 &\cite{ref94}\\
		0.3802 & 86.2  & 14.6 & \cite{ref95} & 1.30  & 168.0 & 17.0  & \cite{ref93} \\
		0.4  & 95.0  & 17.0  & \cite{ref93}  & 1.363  & 160.0 & 33.8 & \cite{ref99} \\
		0.4004 & 79.9  & 11.4 & \cite{ref95} & 1.43  & 177.0 & 18.0  & \cite{ref93}  \\
		0.4247 & 90.4  & 12.8 & \cite{ref95} & 1.53  & 140.0 & 14.0  & \cite{ref93}\\
		0.4497 & 96.3  & 14.4 & \cite{ref95} & 1.75  & 202.0 & 40.0  & \cite{ref93}\\
		0.47  & 89.0  & 49.6  & \cite{ref96} & 1.965  & 186.5 & 50.6 & \cite{ref99}  \\
		\hline
		\multicolumn{8}{|c|}{BAO data} \\
		\hline
		\textbf{$z$} & \textbf{$H(z)$} & \textbf{$\sigma_{H}$} & Ref. & \textbf{$z$} & \textbf{$H(z)$} & \textbf{$\sigma_{H}$} & Ref. \\
		\hline
		0.24 & 79.69 & 2.99 & \cite{ref88}  & 0.57 &96.80 & 3.40  & \cite{ref106}   \\
		0.30 & 81.70 & 6.22  & \cite{ref100} & 0.59 & 98.48 & 3.19  & \cite{ref101}  \\
		0.31 & 78.17 & 6.74  & \cite{ref101}  & 0.6 & 87.90 & 6.10 & \cite{ref104}  \\
		0.34 & 83.17 & 6.74  & \cite{ref88} & 0.61 & 97.30 & 2.10  &  \cite{ref103} \\
		0.35 & 88.1  & 9.45  & \cite{ref102}   & 0.64 & 98.82 & 2.99  & \cite{ref101}  \\
		0.36 & 79.93 & 3.93  & \cite{ref101}  & 0.978 &113.72 & 14.63  & \cite{ref107}  \\
		0.38 & 81.50 & 1.90  & \cite{ref103}  & 1.23 & 131.44 & 12.42  & \cite{ref107}\\
		0.40 & 82.04 & 2.03  & \cite{ref101}& 1.48 & 153.81 & 6.39 &  \cite{ref108}  \\
		0.43 & 86.45 & 3.68  & \cite{ref88}   & 1.526 & 148.11 & 12.71 &  \cite{ref107} \\
		0.44 & 82.60 & 7.80  & \cite{ref104}  & 1.944 & 172.63 & 14.79 &  \cite{ref107}\\
		0.44 & 84.81 & 1.83 &   \cite{ref101}& 2.3 & 224 & 8 & \cite{ref109} \\
		0.48 & 87.79 & 2.03 &  \cite{ref101} & 2.36 & 226.0 & 8&\cite{ref110}  \\
		0.56 & 93.33 & 2.32  &  \cite{ref101}  & 2.4 & 227.8 & 5.61 & \cite{ref111}  \\
		0.57 & 87.60 & 7.80  & \cite{ref105}    & & & & \\
		\hline 
	\end{tabular}
	\caption{The $H(z)$ dataset and associated uncertainties, $\sigma_H$, utilized in our analysis (measured in km s$^{-1}$Mpc$^{-1}$).}
	\label{table1}
\end{table}
\subsection{Observational Hubble Data}
In our study, we use 59 $H(z)$ data points of the Hubble parameter, including 32 points from CC measurements in the redshift range $0.07\le z \le 1.965$ and 27 points from BAO data in the redshift range $0.24\le z \le 2.4$, as described in \cite{ref113}. The BAO dataset comprises 27 points from earlier BAO data, integrating independent datasets such as WiggleZ \cite{ref104}, BOSS DR12 \cite{ref103}, and eBOSS DR16 \cite{ref108, ref114, ref115, ref116, ref117, ref118}. Moresco et al. \cite{ref94, ref95, ref99} contributed 15 data points to the CC dataset, collected using the same procedure. Reference \cite{ref112,ref119} provides further information on the correlation between these points. Our analysis considers the covariance between these points, as described in the open-source program \footnote{\url{https://gitlab.com/mmoresco/CCcovariance/-/tree/master}}. The data is summarized in Table \ref{table1}.
The $32$ CC data points of $H(z)$ are non-correlated hence, we use the following $\chi^{2}$ formula:
\begin{equation}\label{eq43}
	\chi_{CC}^{2}=\sum_{i=1}^{i=32}\frac{[H_{ob}(z_{i})-H_{th}(H_{0}, \xi_{1}, m, \alpha, z_{i})]^{2}}{\sigma_{H(z_{i})}^{2}},
\end{equation}
In this context, $H_{0}, \xi_{1}, m, \alpha$ represent the cosmological parameters that require estimation, while $H_{ob}$ and $H_{th}$ denote the observational and theoretical values of $H(z)$ at $z=z_{i}$, respectively. The $\sigma_{H(z_{i})}$ represents the standard deviations linked to the observed values $H_{ob}$.\\
For $27$ BAO data points of $H(z)$, we use the following expressions of $\chi^{2}$:
\begin{equation}\label{eq44}
	\chi^2_{BAO} = \bigg( H_{th}(H_{0}, \xi_{1}, m, \alpha, z_{i}) - H_{ob}(z_{i})\bigg)^T C_{ij}^{-1} \bigg( H_{th}(H_{0}, \xi_{1}, m, \alpha, z_{j}) - H_{ob}(z_{j})\bigg),
\end{equation}
where $C_{ij}^{-1}$ is the inverse covariance matrix of order $27$.\\
The total $\chi^2$ is then given by
\begin{equation}\label{eq45}
	\chi^2_{CC+BAO} = \chi^2_{CC} + \chi^2_{BAO}.
\end{equation}
\begin{figure}[H]
	\centering
	a.\includegraphics[width=7cm,height=7cm,angle=0]{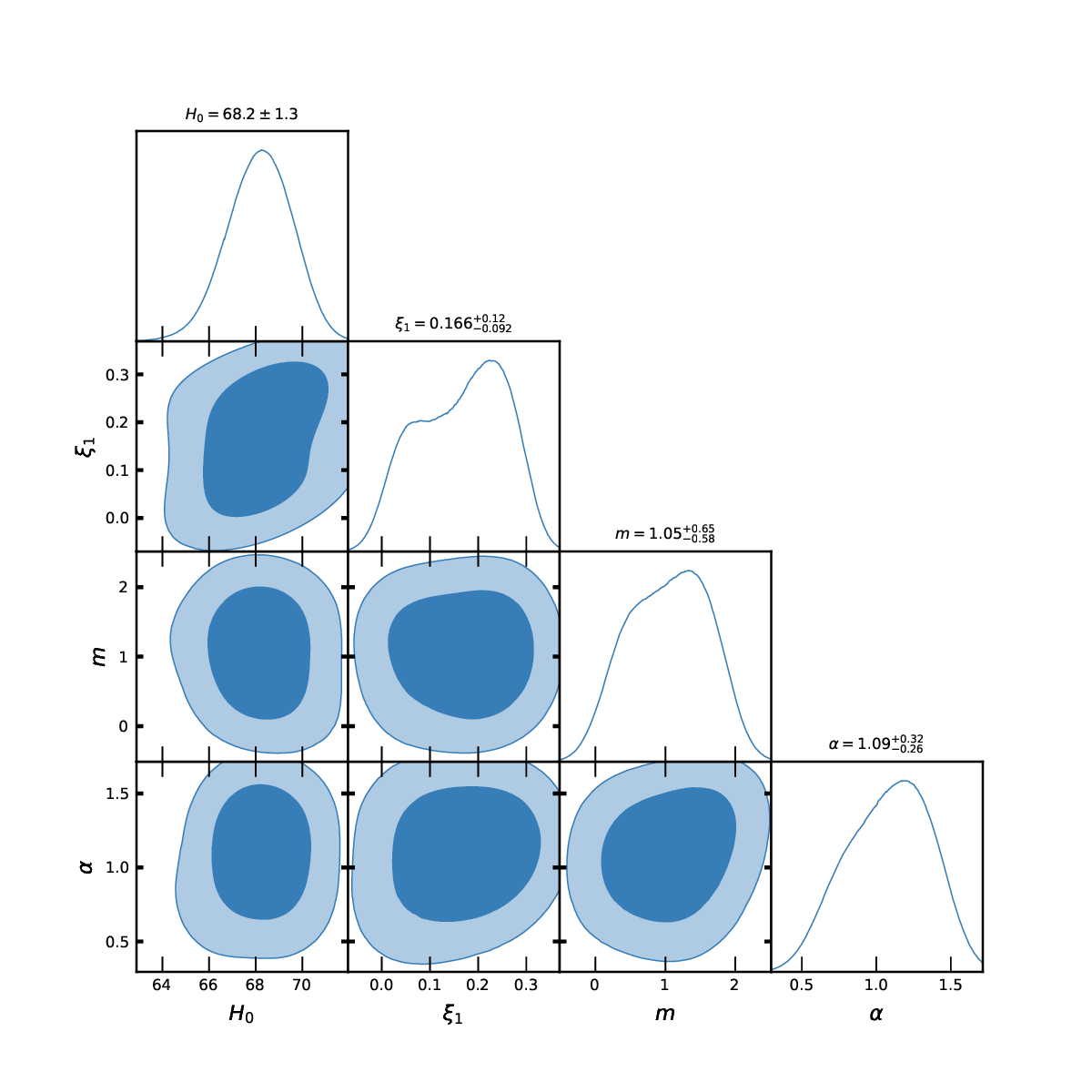}
	b.\includegraphics[width=7cm,height=7cm,angle=0]{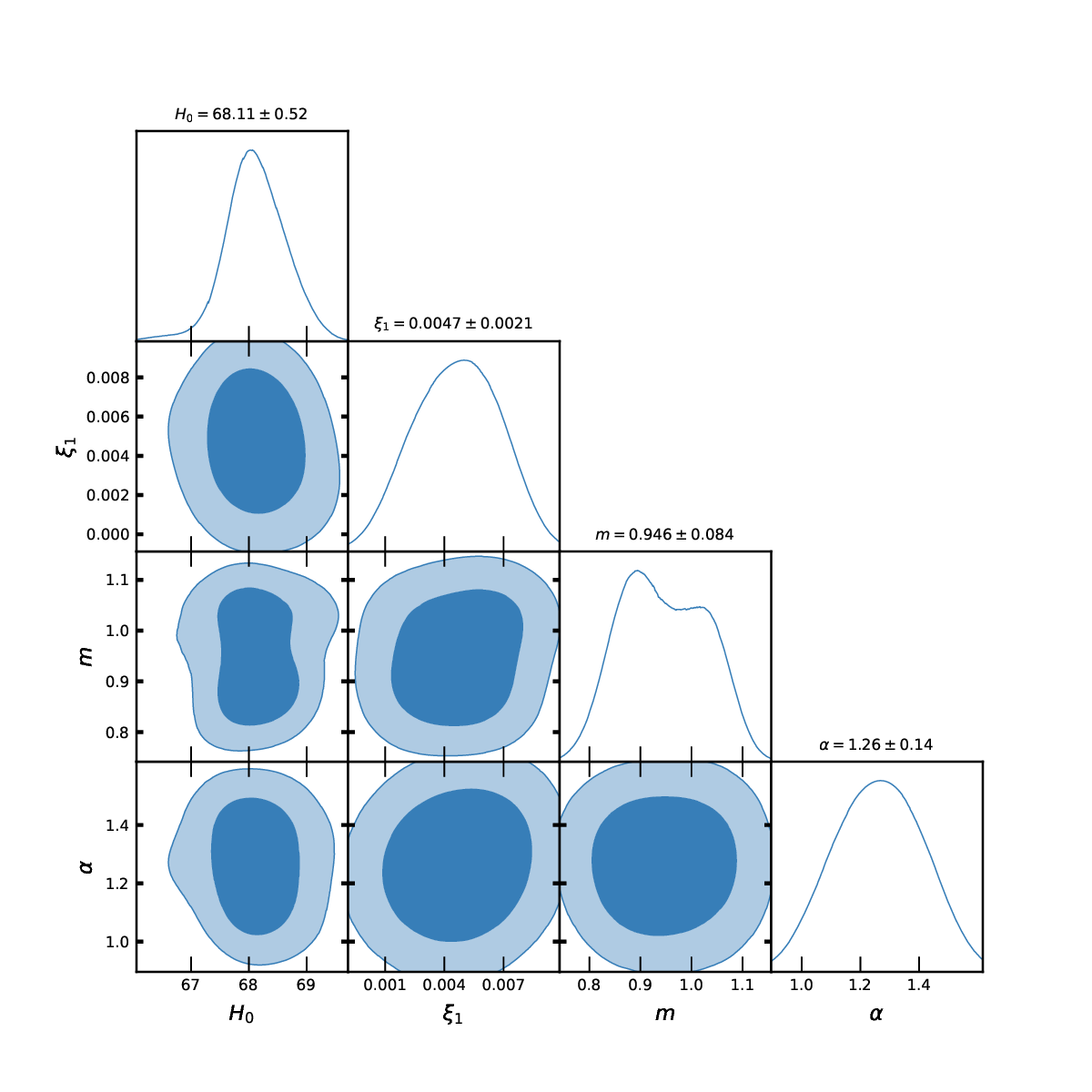}
	\caption{The contour plots of $H_{0}, \xi_{1}, m, \alpha$ at $\sigma1$, $\sigma2$ confidence levels for CC dataset and CC+BAO datasets, respectively.}
\end{figure}
Figure 1a and 1b show the contour plots of $H_{0}, \xi_{1}, m, \alpha$ at a fixed value of arbitrary constant $c_{1}=1.5$ at $\sigma1$, $\sigma2$ confidence levels for CC and CC+BAO datasets, respectively. We have estimated the constrained values of model parameters by applying wide range of priors which mentioned in Table\,\ref{T2}.\\

\subsection{Apparent magnitude}

The data from SNe Ia serves to exemplify the quantification of the expansion rate within the cosmic evolution of the universe, represented through the apparent magnitude $m(z)$. We explored the conceptual framework of apparent magnitude, as articulated in \cite{ref89,ref90,ref120,ref121}.
\begin{equation}\label{eq46}
	m(z)=M+ 5~\log_{10}\left(\frac{D_{L}}{Mpc}\right)+25,
\end{equation}
Here, $M$ represents the absolute magnitude, and the luminosity distance $D_{L}$ is defined in units of length as follows:
\begin{equation}\label{eq47}
	D_{L}=c(1+z)\int^z_0\frac{dz'}{H(z')}.
\end{equation}
The Hubble-free luminosity distance \( d_{L} \) is defined as \( d_{L} \equiv \frac{H_{0}}{c}D_{L} \), indicating a dimensionless quantity. The observable magnitude $m(z)$ can be expressed as
\begin{equation}\label{eq48}
	m(z)=M+5\log_{10}{d_{L}}+5\log_{10}\left(\frac{c/H_{0}}{Mpc}\right)+25.
\end{equation}
A degeneracy between $H_{0}$ and $M$ was observed in the previously described equation, which remains invariant within the $\Lambda$CDM framework \cite{ref120,ref121}.  We will redefine these degenerate parameters for consolidation as follows:
\begin{equation}\label{eq49}
	\mathcal{M}\equiv M+5\log_{10}\left(\frac{c/H_{0}}{Mpc}\right)+25.
\end{equation}
In this context, $\mathcal{M}$ denotes a dimensionless parameter, which can also be formulated as $\mathcal{M}=M-5\log_{10}(h)+42.39$, where $H_{0}=h\times100 \text{ km/s/Mpc}$. The subsequent $\chi^{2}$ formula is employed for the analysis of Pantheon data, as referenced in \cite{ref120}:
\begin{equation}\label{eq50}
	\chi^{2}_{P}=V_{P}^{i}C_{ij}^{-1}V_{P}^{j}.
\end{equation}
The term $V_{P}^{i}$ represents the discrepancy between the observed $m_{ob}(z_{i})$ and the theoretical value $m(\xi_{1}, m, \alpha, \mathcal{M}, z_{i})$ as outlined in equation \eqref{eq48}. Additionally, $C_{ij}^{-1}$ refers to the inverse of the covariance matrix related to the Pantheon sample.\\
We employ the $32$ CC data points for the Hubble parameter in conjunction with the $1048$ Pantheon datasets to derive the joint estimates of model parameters. The $\chi_{CC+P}^{2}$ formula is utilized to perform a joint MCMC analysis of Pantheon and CC data points, facilitating the extraction of combined constraints on the model parameters.
\begin{equation}\label{eq51}
	\chi_{CC+P}^{2}=\chi_{CC}^{2}+\chi^{2}_{P}.
\end{equation}
\begin{table}[H]
	\centering
	\begin{tabular}{|c|c|c|c|c|}
		\hline
		
		Parameter        & Prior              & CC                      & CC$+$BAO          &CC$+$Pantheon\\
		\hline
		$H_{0}$          & $(50, 100)$        & $68.2\pm1.3$            & $68.11\pm0.52$    &$68.4\pm1.6$\\
		$\xi_{1}$        & $(0, 1)$           & $0.166_{-0.092}^{+0.12}$& $0.0047\pm0.0021$ &$0.183_{-0.060}^{+0.098}$\\
		$m$              & $(0, 2)$           & $1.05_{-0.58}^{+0.65}$  & $0.946\pm0.084$   &$1.01\pm0.58$\\
		$\alpha$         & $(0.5, 1.5)$       & $1.09_{-0.26}^{+0.32}$  & $1.26\pm0.14$     &$0.96\pm0.30$\\
		$\mathcal{M}$    & $(23, 24)$         & $-$                     & $-$               &$23.8477\pm0.0051$\\
		$c_{1}$          & Fixed              & $1.5$                   & $1.5$             &$1.5$\\
		$\chi^{2}$       & $-$                & $19.0713$               & $57.3515$               &$1054.8572$\\
		\hline
	\end{tabular}
	\caption{The MCMC estimates.}\label{T2}
\end{table}
\begin{figure}[H]
	\centering
	\includegraphics[width=10cm,height=10cm,angle=0]{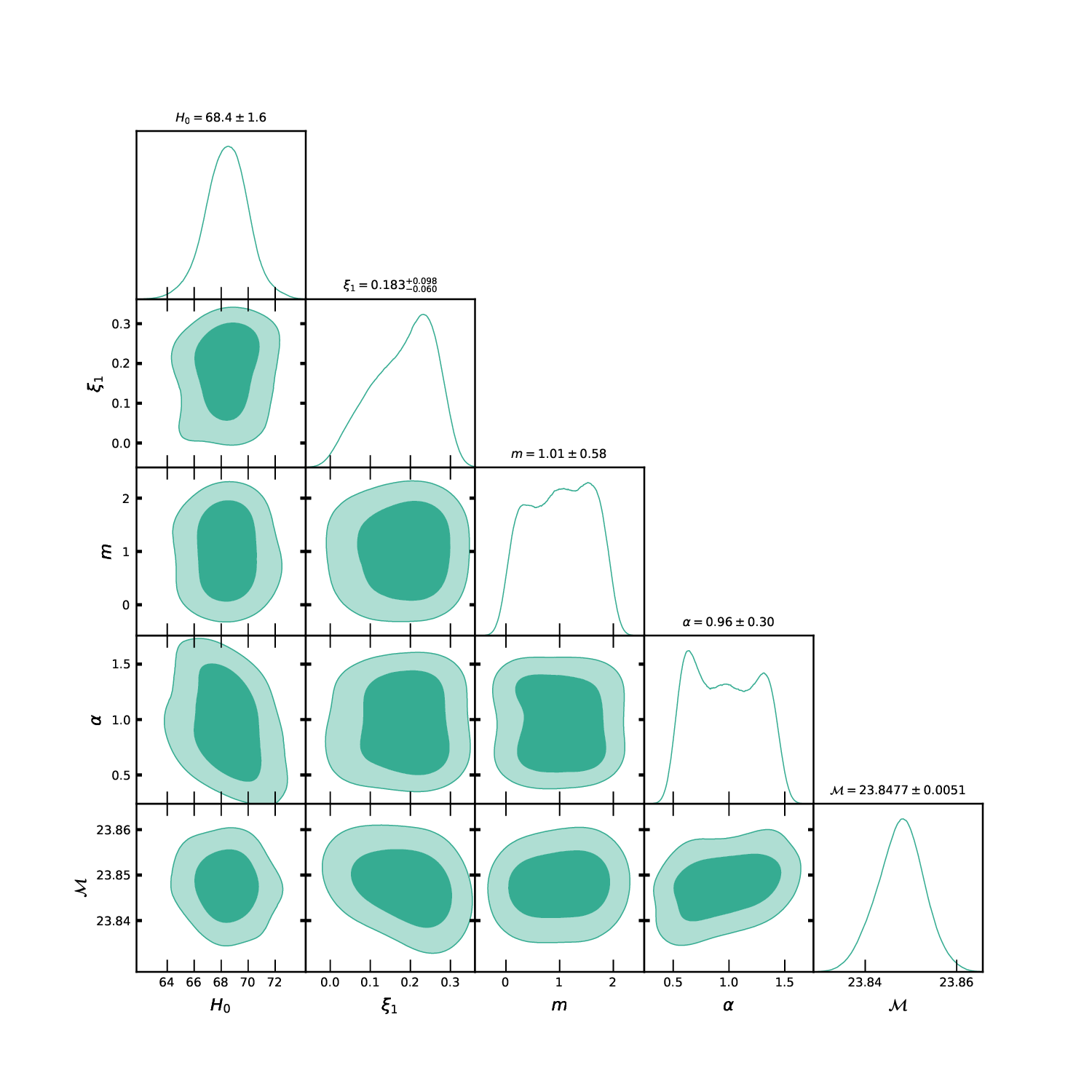}
	\caption{The contour plots of $H_{0}, \xi_{1}, m, \alpha$ and $\mathcal{M}$ at $\sigma1$, $\sigma2$ confidence levels for CC$+$Pantheon datasets.}
\end{figure}
Figure 2 illustrates the contour plots of $H_{0}, \xi_{1}, m, \alpha$, and $\mathcal{M}$ at a constant value of $c_{1}=1.5$, presented at the $\sigma1$ and $\sigma2$ confidence levels for the CC$+$Pantheon dataset.  We estimated the constrained values of model parameters by applying a wide range of priors, as detailed in Table\,\ref{T2}.

\section{Discussion of Results}

This study presents an analytical solution to the field equations in non-linear $f(Q)$ gravity within a locally rotationally symmetric Bianchi type-I spacetime that is filled with viscous fluids. A hyperbolic solution is obtained in relation to the model parameters $\alpha$, $m$, $\xi_{0}$, $\xi_{1}$, $c_{0}$, and $c_{1}$. We conducted MCMC analysis on the CC, CC+BAO and CC$+$Pantheon datasets to derive consistent model parameters values aligned with the observed evolution of the universe. We have examined the behavior of cosmological and physical parameters, including the deceleration parameter $q$, the equation of state parameter $\omega_{v}$, and the skewness parameter $\delta_{v}$, utilizing the estimated values of model parameters across varying redshift $z$. We investigated the cosmic behavior of cosmographic coefficients $H(z), q(z), j(z), s(z), l(z)$ and $m(z)$, as defined in introduction. We conducted an analysis of statefinder parameters and Om diagnostic tests for the classification of dark energy models. We have measured the Hubble constant as $H_{0}=68.2\pm1.3, 68.11\pm0.52, 68.4\pm1.6$ Km/s/Mpc, respectively, along CC, CC+BAO and CC+Pantheon datasets. The model parameters are $\xi_{1}=0.166_{-0.092}^{+0.12}, 0.0047\pm0.0021, 0.183_{-0.060}^{+0.098}$, $m=1.05_{-0.58}^{+0.65}, 0.946\pm0.084, 1.01\pm0.58$, and $\alpha=1.09_{-0.26}^{+0.32}, 1.26\pm0.14, 0.96\pm0.30$, derived from three observational datasets: CC, CC+BAO and CC$+$Pantheon, respectively. The constrained value of the dimensionless parameter $\mathcal{M}$ has been estimated as $23.8477\pm0.0051$, which varies upon the theoretical models (see \cite{ref122,ref123,ref124,ref125,ref126,ref127,ref128,ref129}).\\
\begin{figure}[H]
	\centering
	a.\includegraphics[width=8cm,height=7cm,angle=0]{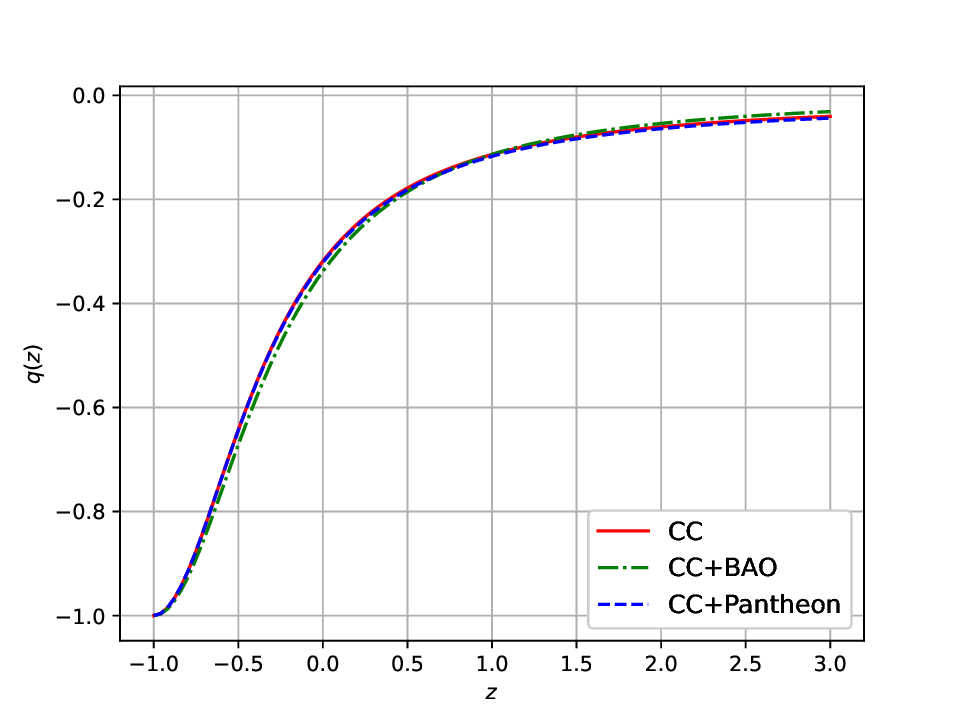}
	b.\includegraphics[width=8cm,height=7cm,angle=0]{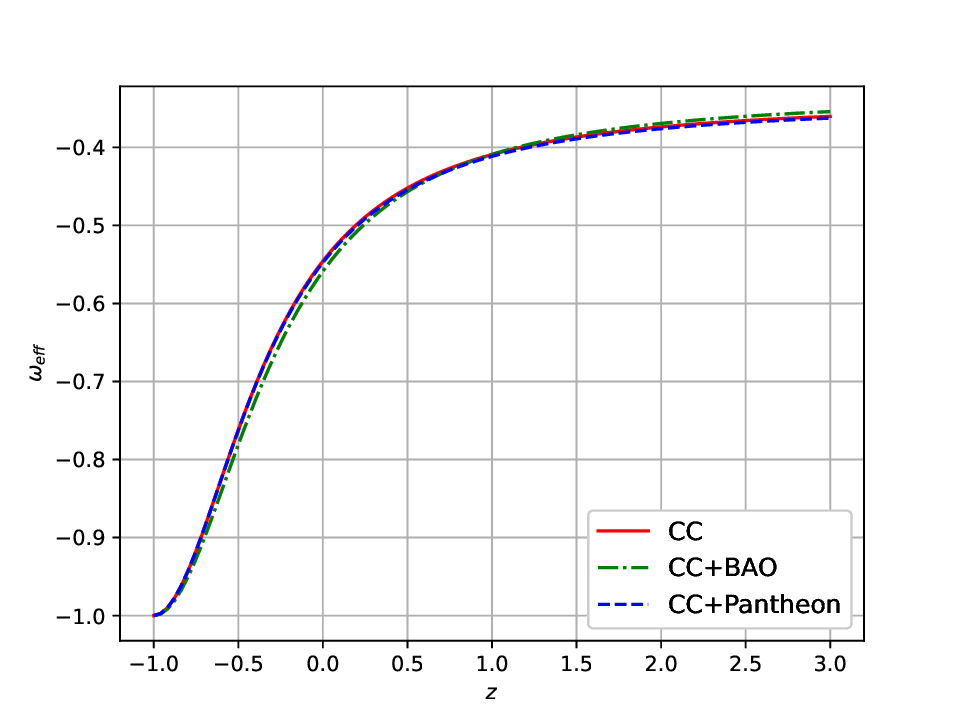}
	\caption{The plots of deceleration parameter $q(z)$ and effective EoS parameter $\omega_{eff}$ versus $z$, respectively.}
\end{figure}
The dimensionless parameter $q$ characterizes the phase of the expanding universe; a positive value indicates a decelerating phase, whereas a negative value signifies an accelerating phase of expansion. The deceleration parameter $q(z)$ as a function of $z$ is presented in Equation \eqref{eq39}. Figure 3a illustrates the variation of $q(z)$ with respect to redshift $z$. From Figure 3a one can observe that the deceleration parameter values range from $-1$ to $0$ across the redshift $z$. The values of $q_{0}$ are determined to be $-0.3185$, $-0.3380$, and $-0.3211$ based on three observational datasets, CC, CC+BAO and CC$+$Pantheon, respectively. This indicates that the universe's evolution in our model is continuously accelerating, aligning with recent observations. As \( z \to \infty \), it follows that \( q \to -1 + \frac{36\alpha(2m+1)}{36\alpha(2m+1) + \xi_{1}(m+2)^{2}} \) which depicts the dependency of accelerating phase of the universe on model parameters $\alpha, m$ and $\xi_{1}$. The effective EoS parameter for the model is defined as $\omega_{eff}=\frac{2q-1}{3}$, $q$ is the deceleration parameter given in Eq.\,\eqref{eq39}. Figure 3b depicts the evolution of $\omega_{eff}$ over $z$ and one can observed that the whole evolution of effective EoS parameter as $-1\le\omega_{eff}<-\frac{1}{3}$ that is compatible with ever accelerating model. We have estimated the present values of $\omega_{eff}=-0.5456, -0.5587, -0.5474$, respectively, along three datasets CC, CC+BAO and CC+Pantheon.\\

The equations \eqref{eq40} and \eqref{eq42} provide the values for the equation of state parameter (EoS) $\omega_{v}$ and the skewness parameter $\delta_{v}$ in the context of a bulk viscosity fluid within an anisotropic spacetime universe. Figures 4a and 4b illustrate the variation of these parameters over redshift $z$. Figures 4a and 4b illustrate that the parameters $\omega_{v}$ and $\delta_{v}$ increase with rising redshift $z$. A universe characterized by a viscous fluid behaves similarly to a potential dark energy candidate. In our model we have measured the present value of EoS $\omega_{v}$ as $-0.4507$, $-0.4755$ and $-0.4561$, along with the current values of $\delta_{v}$ as $-0.0064$, $-0.0076$ and $-0.0013$, derived from three distinct observational datasets. Figure 4a indicates that as \( z \to -1 \), \( \omega_{v} \) approaches $-0.9871$, $-1.0149$ and $-0.9973$, respectively, across three observational datasets. Figure 4b illustrates that the skewness parameter $\delta_{v}$ increases with rising redshift $z$. The values of $\delta_{v}$ vary within the interval $(-0.02< \delta_{v}< 0.03)$, consistent with the property of skewness. Furthermore, it is observed that as \( z \to \infty \), \( \delta_{v} \to 0 \), and in the late-time universe, it is different from zero. Consequently, it can be stated that the strength of the viscous force diminishes over time, leading to the accelerating expansion in the evolution of the universe.\\
\begin{figure}[H]
	\centering
	a.\includegraphics[width=8cm,height=7cm,angle=0]{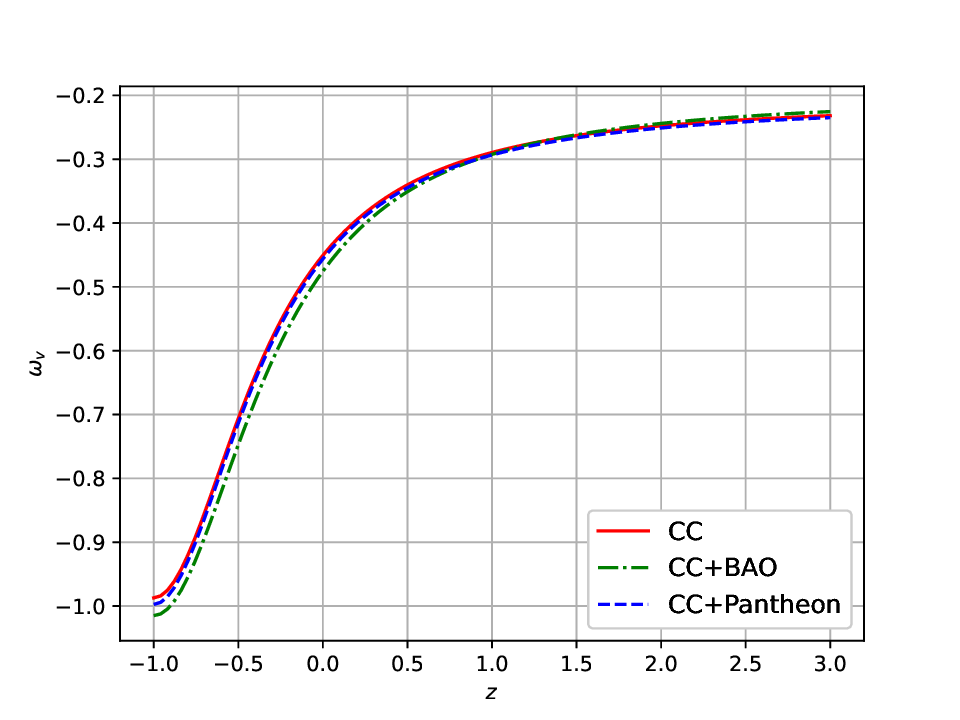}
	b.\includegraphics[width=8cm,height=7cm,angle=0]{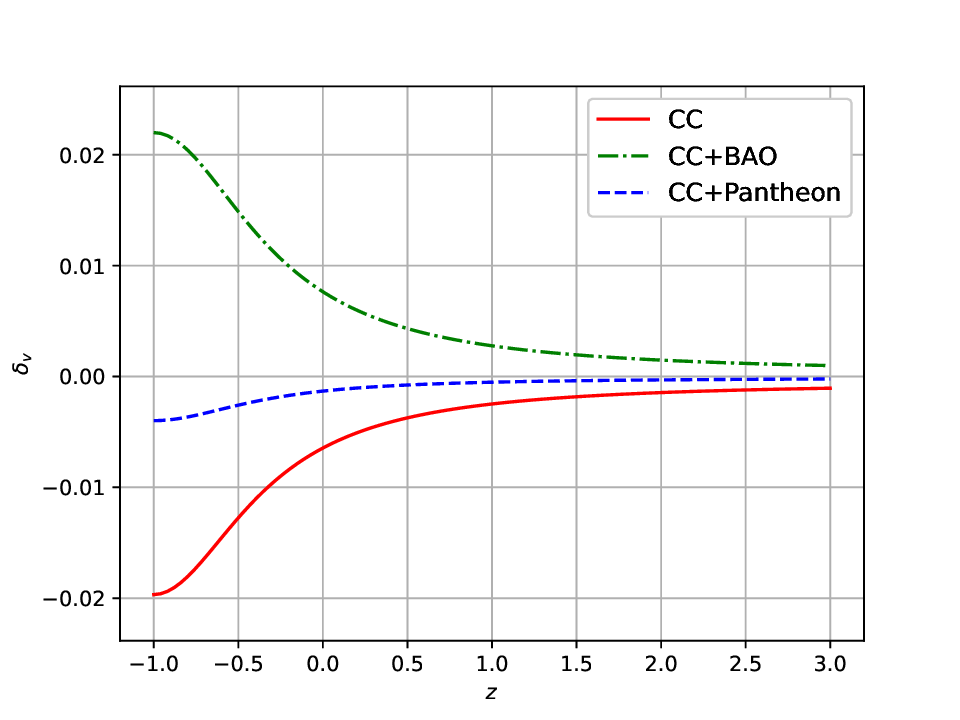}
	\caption{The plots of EoS parameter $\omega_{v}$ and skewness parameter $\delta_{v}$ versus $z$, respectively.}
\end{figure}

\subsection{Cosmographic Analysis}
The cosmological principle specifies a scale factor as the only degree of freedom that rules the universe. By expanding the current Taylor series of $a(t)$ around present time, we may construct the cosmographic series coefficients such as Hubble parameter $(H)$, deceleration parameter $(q)$, jerk $(j)$, snap $(s)$, lerk $(l)$, and max-out $(m)$ as presented in \cite{ref66}.
\begin{equation}\label{eq52}
	H=\frac{1}{a}\frac{da}{dt},~~q=-\frac{1}{aH^{2}}\frac{d^{2}a}{dt^{2}},~~j=\frac{1}{aH^{3}}\frac{d^{3}a}{dt^{3}}
\end{equation}
and
\begin{equation}\label{eq53}
	s=\frac{1}{aH^{4}}\frac{d^{4}a}{dt^{4}},~~l=\frac{1}{aH^{5}}\frac{d^{5}a}{dt^{5}},~~m=\frac{1}{aH^{6}}\frac{d^{6}a}{dt^{6}}
\end{equation}
\begin{figure}[H]
	\centering
	a.\includegraphics[width=8cm,height=7cm,angle=0]{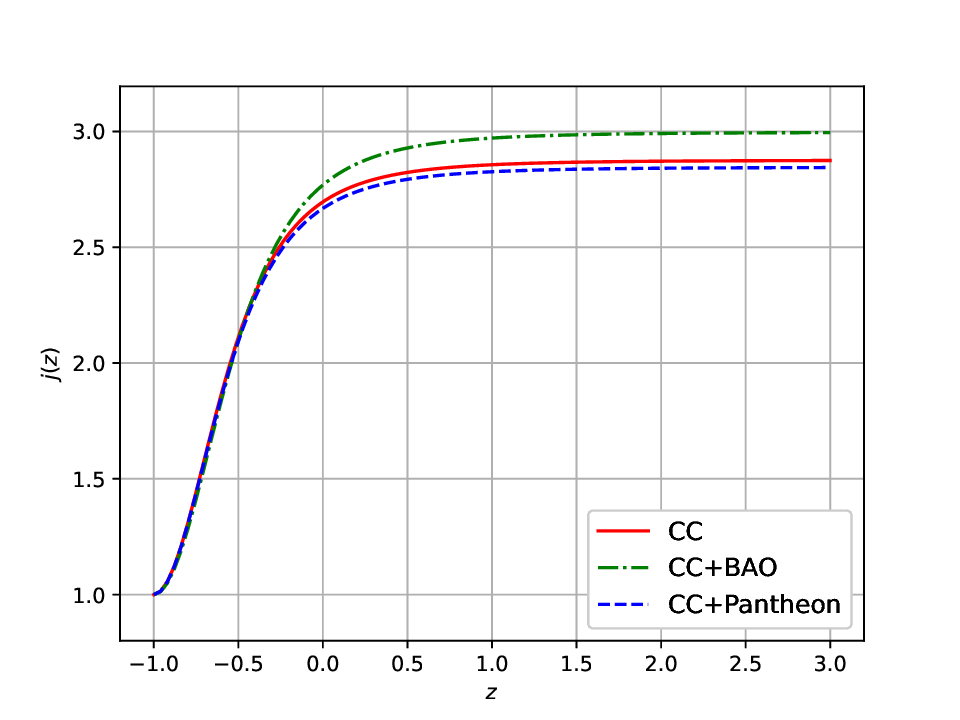}
	b.\includegraphics[width=8cm,height=7cm,angle=0]{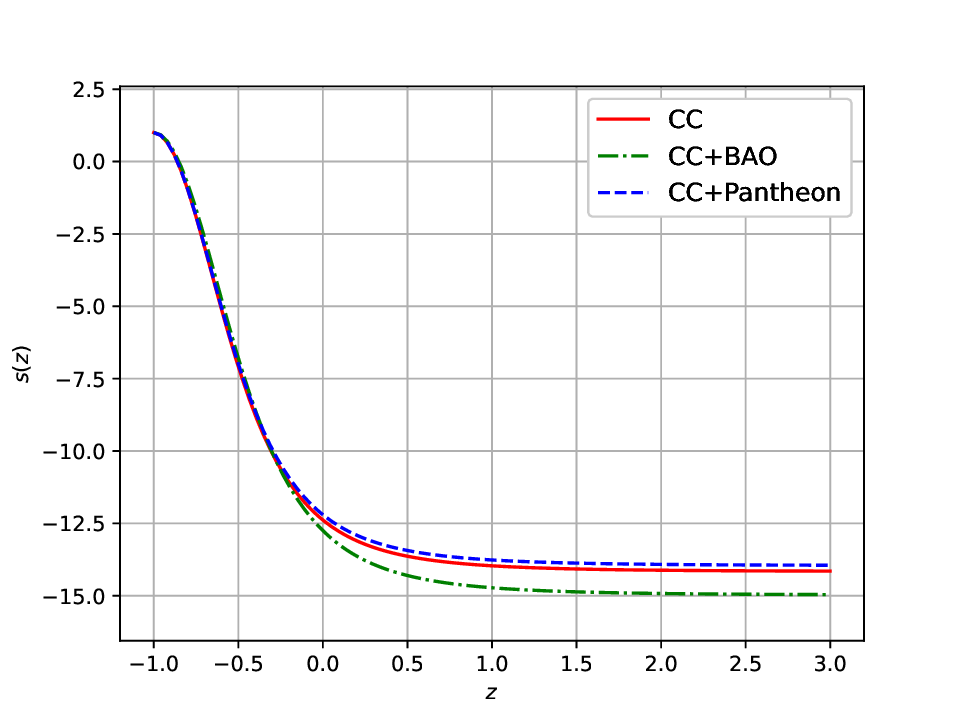}
	\caption{The plots of jerk parameter $j(z)$ and snap parameter $s(z)$ over $z$, respectively.}
\end{figure}
Using these variables, researchers investigate the dynamics of the universe in its later stages. In order to ascertain the physical properties of the coefficients, the form of the Hubble expansion might be utilized. To be more specific, the sign of the parameter $q$ tells us whether the universe is speeding up or slowing over. Positive values of $j$ indicate the occurrence of transitional intervals when the expansion of the universe transits from decelerating to accelerating or accelerating to decelerating, and the sign of $j$ determines how the dynamics of the universe adapt to new circumstances. Additionally, we need to know the value of $s$ in order to differentiate between the ever-evolving theories of dark energy and the behavior of cosmological constants.\\
Using the scale-factor \eqref{eq37} in \eqref{eq52} and \eqref{eq53}, we have derived the cosmographic series coefficients $q, j, s, l, m$ as
\begin{eqnarray}
	q(t)&=&-1+\frac{1}{n}\,sech^{2}(k_{1}t+c_{0}),\label{eq54}\\
	j(t)&=&\frac{(n-1)(n-2)}{n^{2}}+\frac{3n-2}{n^{2}}\,\tanh^{2}(k_{1}t+c_{0}),\label{eq55}\\
	s(t)&=&\frac{(n-1)(n-2)(n-3)}{n^{3}}+\frac{2(n-1)(3n-4)}{n^{3}}\,\tanh^{2}(k_{1}t+c_{0})+\frac{3n-2}{n^{3}}\,\tanh^{4}(k_{1}t+c_{0}),\label{eq56}\\
	l(t)&=&\frac{(n-1)(n-2)(n-3)(n-4)}{n^{4}}+\frac{2(n-1)(n-2)^{2}}{n^{4}}\,\tanh^{2}(k_{1}t+c_{0})+\frac{15n^{2}-30n+16}{n^{4}}\,\tanh^{4}(k_{1}t+c_{0}),\label{eq57}\\
	m(t)&=&\frac{(n-1)(n-2)(n-3)(n-4)(n-5)}{n^{5}}+\frac{(n-1)(n-2)(n-3)(7n-24)}{n^{5}}\,\tanh^{2}(k_{1}t+c_{0})\nonumber\\
	&&+\frac{(n-1)(21n^{2}-54n+40)}{n^{5}}\,\tanh^{4}(k_{1}t+c_{0})+\frac{15n^{2}-30n+16}{n^{5}}\,\tanh^{6}(k_{1}t+c_{0}).\label{eq58}
\end{eqnarray}
The characteristics of the first two cosmographic parameters, $H$ and $q$, have been addressed previously in this section. The subsequent cosmographic coefficient is the jerk parameter $j(t)$, as defined by Eq.\,\eqref{eq55}. Its variation with respect to redshift $z$ is illustrated in Figure 5a. The jerk parameter consistently yields a positive value $(j>0)$, suggesting the presence of a transition period during which the universe alters its expansion phase. The current estimation of the jerk parameter in our model is $j_{0}=2.6957, 2.7689, 2.6677$ across three data sets, respectively, with a variation range of $1\le j \le 3$ over the redshift interval of $-1\leq z\leq3$ \cite{ref66}. The subsequent cosmographic coefficient is the snap parameter $s$, as defined by Eq.\,\eqref{eq56}, and its variation with respect to $z$ is illustrated in Figure 5b. The snap parameter indicates the behavior of the dark energy term or cosmological constant within the model. The calculated present value of the snap parameter is $s_{0}=-12.3857, -12.7508, -12.2031$ for the three data sets: CC, CC+BAO, and CC+Pantheon, respectively. The additional cosmographic coefficients are lerk $l(t)$ and max-out $m(t)$, with their expressions provided by Eqs.\,\eqref{eq57} and \eqref{eq58}, respectively. Their behaviors as a function of $z$ are illustrated in Figures 6a and 6b, respectively. The estimated present values are $l_{0}=95.1577, 101.7009, 93.4549$ and $m_{0}=-761.0965, -806.4480, -744.5332$ across three data sets, respectively. These values fluctuate with cosmic redshift $z$ within the ranges of $(90, 145)$ and $(-1000, 200)$, respectively. Consequently, it becomes evident that as $t\to\infty$ (or $z\to-1$), the set $\{q, j, s\}$ approaches $\{-1, 1, 1\}$, highlighting a favorable aspect of our derived model \cite{ref66,ref67,ref68,ref69,ref70,ref71,ref72,ref73,ref74}.\\

\begin{figure}[H]
	\centering
	a.\includegraphics[width=8cm,height=7cm,angle=0]{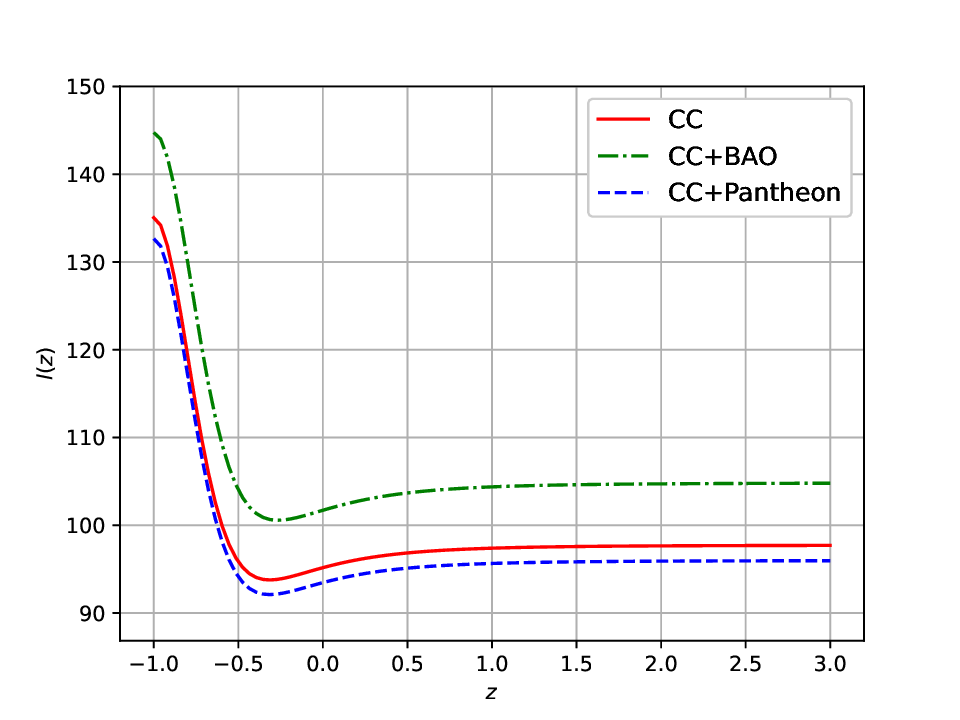}
	b.\includegraphics[width=8cm,height=7cm,angle=0]{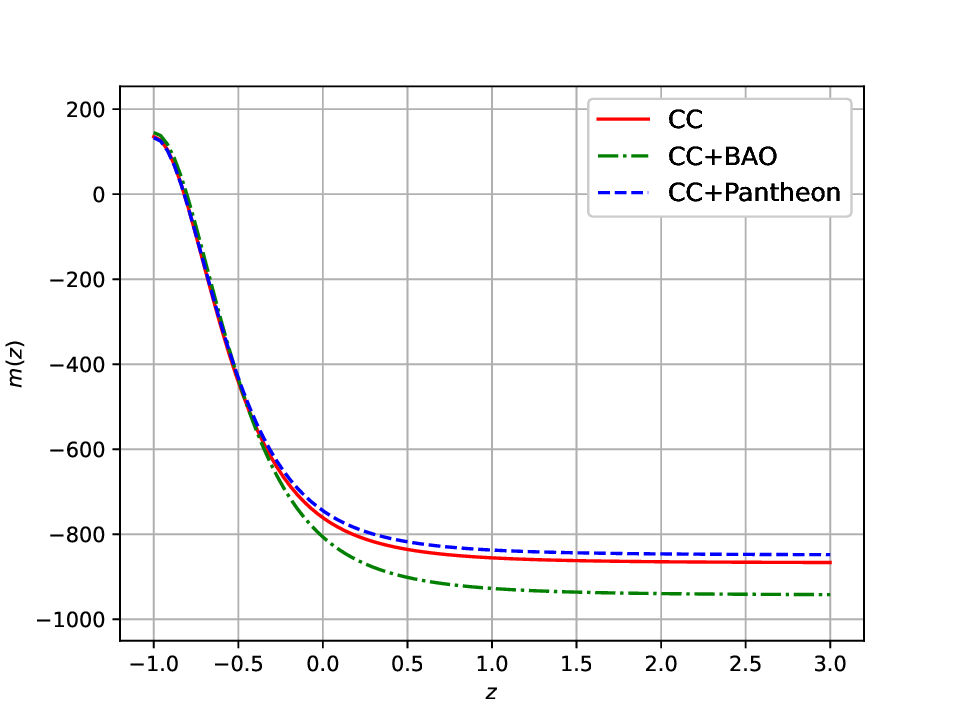}
	\caption{The plots of lerk parameter $l(z)$ and max-out parameter $m(z)$ over $z$, respectively.}
\end{figure}

\subsection{Age of the present universe}
We define the age of universe as follows:
\begin{equation}\label{eq59}
t_{0}-t=-\int_{t_{0}}^{t}dt=\int_{0}^{z}\frac{dz'}{(1+z')H(z')}
\end{equation}
Using \eqref{eq38} in \eqref{eq59} and integrating, we get
\begin{equation}\label{eq60}
t_{0}-t=\frac{n\sqrt{1+c_{1}^{\frac{2}{n}}}}{H_{0}}\left[\tanh^{-1}\sqrt{1+c_{1}^{\frac{2}{n}}}-\tanh^{-1}\sqrt{1+[c_{1}(1+z)]^{\frac{2}{n}}}\right]\times978\,\,\,\,\text{(in Giga Years)}
\end{equation}
Figure 7 illustrates the relationship between the cosmic age of the universe, represented as $t_{0}-t$, and redshift $z$. The present age of the universe has been determined to be $t_{0}=13.82$, $13.89$ and $13.81$ Gyrs, based on three observational datasets, CC, CC+BAO and CC$+$Pantheon, respectively. These findings align with recent observed values reported in various studies.
\begin{figure}[H]
	\centering
	\includegraphics[width=9cm,height=7cm,angle=0]{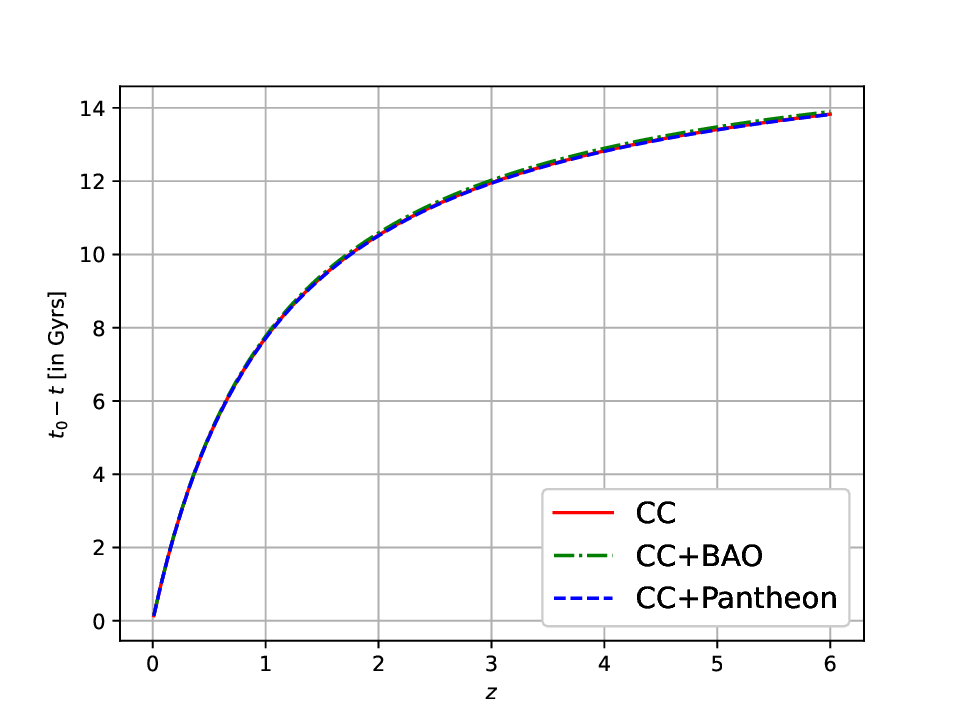}
	\caption{The evolution of cosmic age of the universe versus $z$.}
\end{figure}
\subsection{Statefinder Analysis}
In cosmology, two geometrical parameters are recognized: the Hubble parameter \( H = \frac{\dot{a}}{a} \) and the deceleration parameter \( q = -\frac{a\ddot{a}}{\dot{a}^{2}} \), where \( a(t) \) represents the average scale factor. These parameters characterize the history of the universe. Additional geometrical parameters, known as statefinder diagnostics, have been proposed in \cite{ref130} to represent the geometric evolution of various stages of dark energy models \cite{ref130,ref131,ref132}. The statefinder parameters $r$ and $S$ are defined in relation to the average scale factor $a(t)$ as follows:
\begin{equation}\label{eq61}
  r=\frac{\dddot{a}}{aH^{3}},~~~~~~~~S=\frac{r-1}{3(q-\frac{1}{2})}
\end{equation}
In the present model, we derive the expression for $r$, as
\begin{equation}\label{eq62}
  r=1-\frac{3n-2}{n^{2}}\,sech^{2}(k_{1}t+c_{0})
\end{equation}
And the expression for $S$ as
\begin{equation}\label{eq63}
  S=\frac{2(3n-2)\,sech^{2}(k_{1}t+c_{0})}{3n[3n-2\,sech^{2}(k_{1}t+c_{0})]}
\end{equation}
\begin{figure}[H]
	\centering
	a.\includegraphics[width=8cm,height=6cm,angle=0]{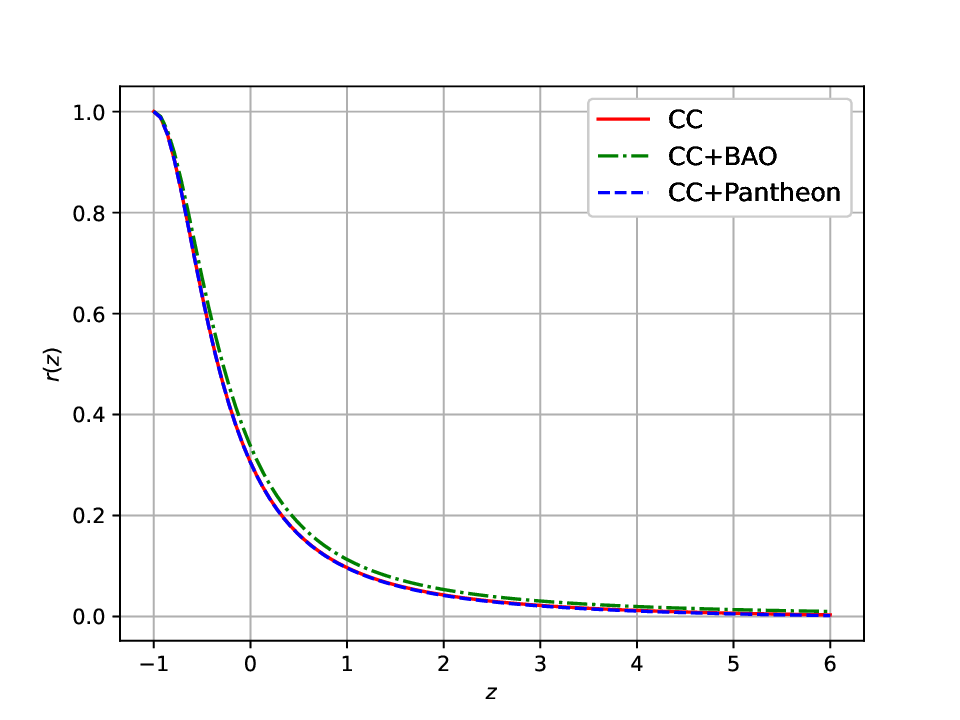}
    b.\includegraphics[width=8cm,height=6cm,angle=0]{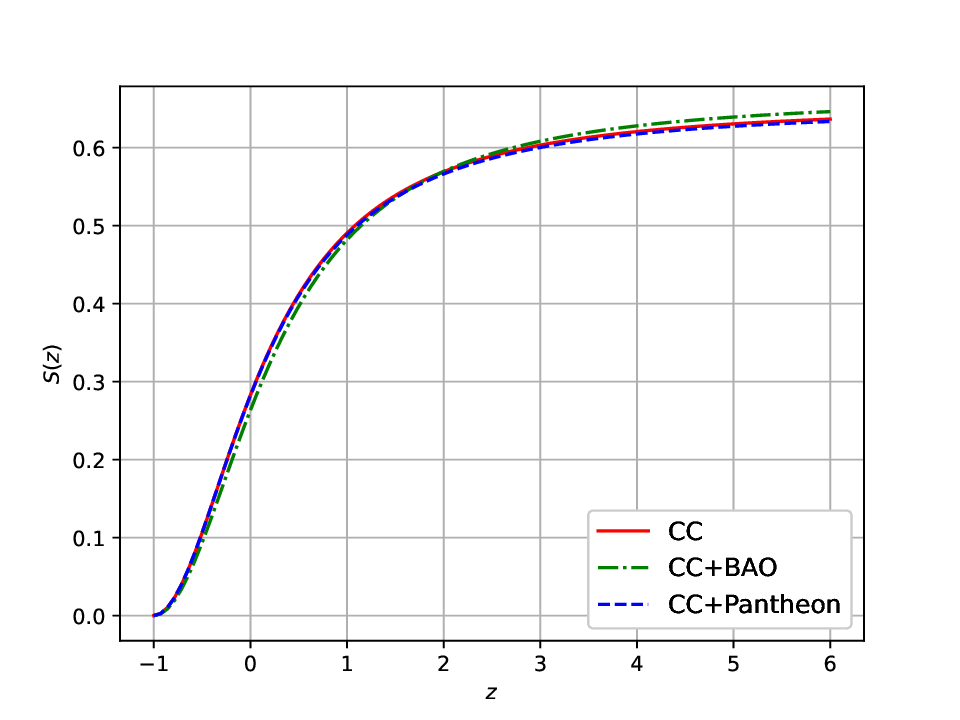}
	\caption{The variations of statefinder parameters $r(z)$ and $s(z)$ versus $z$, respectively.}
\end{figure}
\begin{figure}[H]
	\centering
	a.\includegraphics[width=8cm,height=6cm,angle=0]{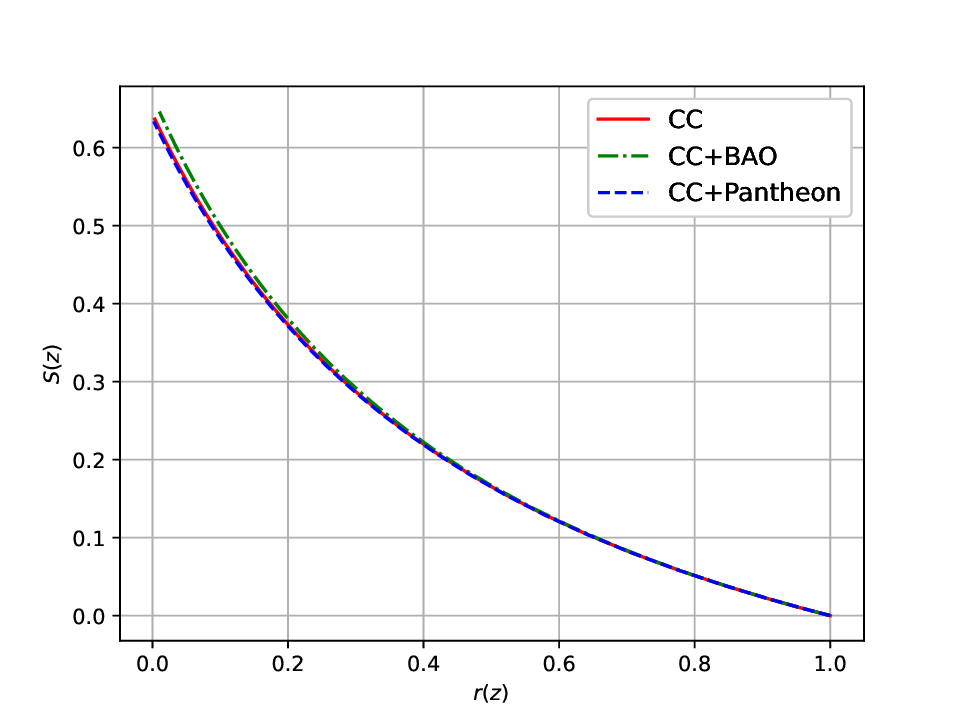}
    b.\includegraphics[width=8cm,height=6cm,angle=0]{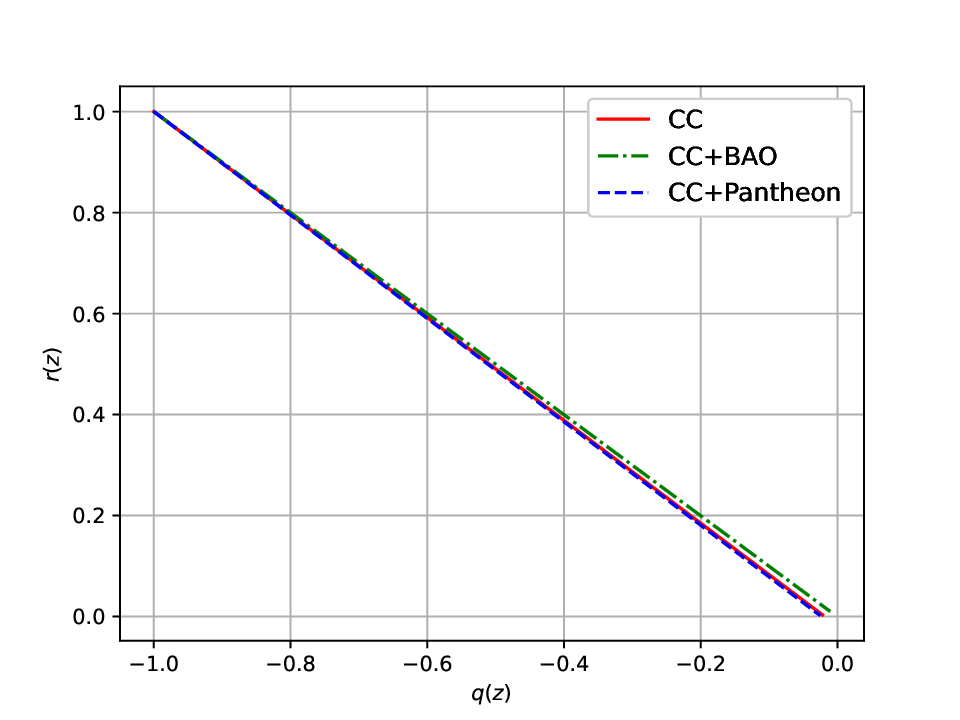}
	\caption{The variations of $s(z)$ versus $r(z)$, and $r(z)$ versus $q(z)$, respectively.}
\end{figure}
The behaviors of $r$ and $S$ over $z$ are illustrated in figures 8a and 8b, respectively. The present values measured are $r_{0}=\{0.3055, 0.3377, 0.3050\}$ and $S_{0}=\{0.2828, 0.2633, 0.2821\}$ for the three data sets, respectively. As \( z \to -1 \), it follows that \( r \to 1 \) and \( s \to 0 \).  Figures 9a and 9b illustrate the plots of \(S-r\) and \(r-q\), respectively. The variation of \((S,r)\) indicates different dark energy models \cite{ref130, ref131, ref132}; for instance, the point \((S,r)=(0,1)\) corresponds to the \(\Lambda\)CDM, a flat FLRW model. Figure 9b indicates that the current values are $(r_{0},q_{0})=(0.3055, -0.3185)$, $(0.3377, -0.3380)$ and $(0.3050, -0.3211)$ for the three data sets, suggesting that our present universe is either matter-dominated or dark energy-dominated.

\subsection{Om diagnostic}

The behavior of the Om diagnostic function \cite{ref133} allows for the categorization of theories regarding cosmic dark energy. The Om diagnostic function for a spatially homogeneous universe is defined as follows.
\begin{equation}\label{eq64}
	Om(z)=\frac{\left(\frac{H(z)}{H_{0}}\right)^{2}-1}{(1+z)^{3}-1},
\end{equation}
Here, $H_{0}$ represents the present value of the Hubble parameter, while $H(z)$ denotes the Hubble parameter as defined in Eq.\,\eqref{eq38}. A positive slope of \( Om(z) \) indicates phantom motion, whereas a negative slope signifies quintessence motion. The $\Lambda$CDM model is characterized by the constant $Om(z)$.\\
\begin{figure}[H]
	\centering
	\includegraphics[width=10cm,height=8cm,angle=0]{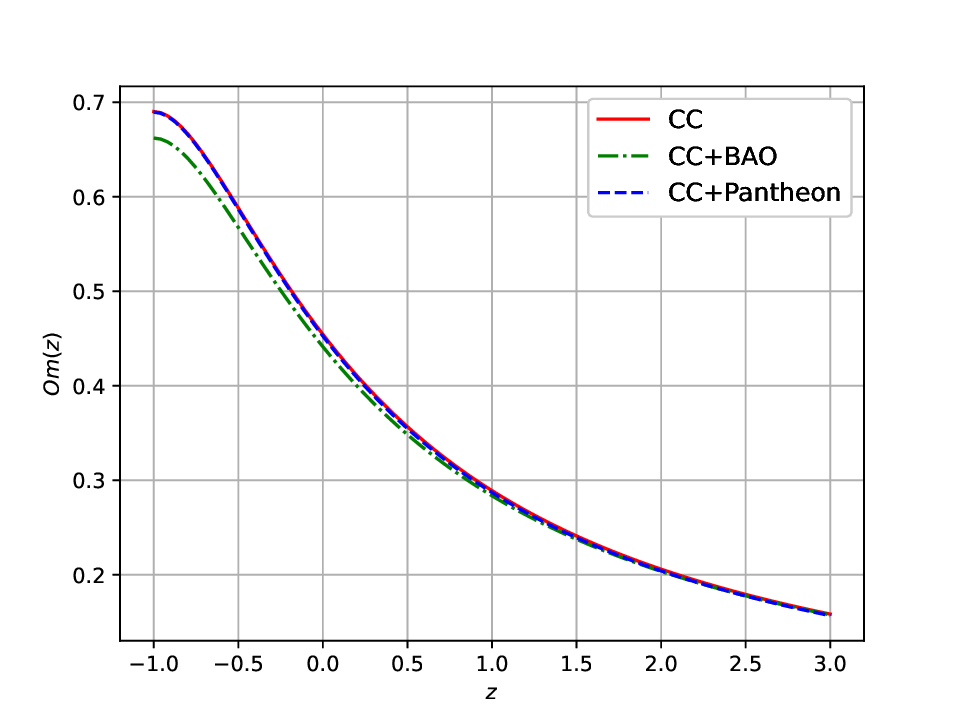}
	\caption{The variation of $Om(z)$ versus $z$.}
\end{figure}
Figure 10 illustrates the behavior of the $Om(z)$ function over $z$ for the model we derived. Figure 10 indicates that the slope of the $Om(z)$ curve is negative, suggesting that our universe model exhibits characteristics similar to those of a quintessence dark energy model. Furthermore, it is evident that in the late-time future, the value of $Om(z)$ approaches a constant, indicating that our derived model converges to the $\Lambda$CDM stage in late-time future.

\subsection{Other Physical Parameters}

In this section, we derived additional physical parameters, including the expansion scalar $\theta$, shear scalar $\sigma$, and anisotropy parameter $\Delta$, as outlined below:
\begin{equation}\label{eq65}
	\theta(t)=3H
\end{equation}
\begin{equation}\label{eq66}
	\sigma^{2}(t)=3\left(\frac{m-1}{m+2}\right)^{2}H^{2}
\end{equation}
\begin{equation}\label{eq67}
	\Delta=2\left(\frac{m-1}{m+2}\right)^{2}
\end{equation}
From Eqs.\,\eqref{eq65} and \eqref{eq66}, we have estimated the values of the ratio $\sigma/\theta\approx0.0095, -0.010582, 0.0019$, respectively, along three datasets CC, CC+BAO and CC$+$Pantheon which is about $10^{-3}$ strength. From Eq.\,\eqref{eq55}, we estimated the value of $\sigma/H\approx0.02839, -0.031748, 0.00574$, along CC, CC+BAO and CC$+$Pantheon datasets while using Eq.\,\eqref{eq67}, we have estimated the values of anisotropy parameter $\Delta\approx0.000537, 0.000671, 0.000022$ which indicates that our derived model approaches to a flat, homogeneous and isotropic $\Lambda$CDM model.

\section{Conclusions}

        In this study, we have examined a locally rotationally symmetric (LRS) Bianchi type-I cosmological model within the framework of non-linear $f(Q)$ gravity, incorporating observational constraints. The modified Einstein's field equations were solved using a viscous fluid source, resulting in a hyperbolic solution expressed as $a(t)=c_{1}[\sinh(k_{1}t+c_{0})]^{n}$. Initially, we establish observational constraints on model parameters through MCMC analysis of the cosmic chronometer (CC), BAO and Pantheon datasets. We have measured the Hubble constant as $H_{0}=68.2\pm1.3, 68.11\pm0.52, 68.4\pm1.6$ Km/s/Mpc, respectively, along CC, CC+BAO and CC+Pantheon datasets. The model parameters are $\xi_{1}=0.166_{-0.092}^{+0.12}, 0.0047\pm0.0021, 0.183_{-0.060}^{+0.098}$, $m=1.05_{-0.58}^{+0.65}, 0.946\pm0.084, 1.01\pm0.58$, and $\alpha=1.09_{-0.26}^{+0.32}, 1.26\pm0.14, 0.96\pm0.30$, derived from three observational datasets: CC, CC+BAO and CC$+$Pantheon, respectively. We have studied the behavior of cosmological parameters, including the Hubble parameter $H$, the deceleration parameter $q$, and the equation of state (EoS) parameter $\omega_{v}$, utilizing the estimated values of model parameters alongside the skewness parameter $\delta_{v}$ for the viscous fluid. An accelerating universe model is presented with current deceleration parameter value of $q_{0}=-0.3185, -0.3380$ and $q_{0}=-0.3211$, alongside equation of state parameters $\omega_{v}=-0.4507, 0.4755$ and $\omega_{v}=-0.4561$, derived from three distinct observational datasets. We investigated the behavior of the skewness parameter $\delta_{v}$ across $z$ and estimated its present value as $\delta_{v}=-0.00645, -0.0076$ and $\delta_{v}=-0.00131$ for three physically consistent datasets, respectively. We have estimated the present values of $\omega_{eff}=-0.5456, -0.5587, -0.5474$, respectively, along three datasets CC, CC+BAO and CC+Pantheon. We have studied the behavior cosmographic coefficients $q, j, s, l, m$ that reveals the quintessence dark energy property and approaching to $\Lambda$CDM model at late-time universe. We have examined Om diagnostic and statefinder analysis to categorize dark energy models. The model presented is a quintessence-accelerating framework incorporating bulk-viscosity fluid, converging towards the $\Lambda$CDM paradigm in late-time phases. The current age of the universe is estimated to be approximately $13.8$ billion years. Our investigation of the physical and kinematic parameters revealed that the ratios $\sigma/\theta\approx0.0095, -0.010582, 0.0019$ and $\sigma/H\approx0.02839, -0.031748, 0.00574$ exhibited similarity in all CC, CC+BAO and CC+Pantheon datasets. The anisotropy parameter values were $\Delta\approx0.000537, 0.000671, 0.000022$, indicating that our model closely resembles a flat, homogeneous, and isotropic $\Lambda$CDM model. A late-time accelerating feature is observed in a non-linear $f(Q)$ theory with a viscous fluid source, without the necessity of incorporating a $\Lambda$ cosmological constant term.

\section*{Acknowledgments}

The author is thankful to the renowned reviewer and editor for their valuable suggestions to improve the quality of this manuscript. The author is thankful to IUCAA Center for Astronomy Research and Development (ICARD), CCASS, GLA University, Mathura, India for providing facilities and support where part of this work is carried out.

\section{Data Availability Statement}
No data associated in the manuscript.

\section{Declarations}
\subsection*{Funding and/or Conflicts of interests/Competing interests}
The author of this article has no conflict of interests. The author has no competing interests to declare that are relevant to the content of this article. The author did not receive support from any organization for the submitted work.

\end{document}